\documentclass[%
 sor,
groupedaddress,
 jor,
 amsmath,amssymb,
preprint,%
 reprint,%
author-year,%
]{revtex4-2}
\usepackage{graphicx}
\usepackage{dcolumn}
\usepackage{bm}
\usepackage{fancyhdr}
\usepackage[normalem]{ulem}
\usepackage{amsmath}
\usepackage{textcomp, gensymb}
\usepackage{url}
\usepackage{amsmath,amssymb,stmaryrd}   
\usepackage{listings}
\usepackage{tcolorbox}
\usepackage{blindtext}  
\usepackage{tikz,lipsum,lmodern}
\usepackage{xcolor}     
\usepackage{lipsum}     
\usepackage{ntheorem}   
\usepackage{mdframed}   
\usepackage{gensymb}
\theoremstyle{break}
\theoremheaderfont{\bf}
\theoremstyle{nonumberplain}
\newmdtheoremenv[%
linecolor=gray,leftmargin=60,%
rightmargin=30,
backgroundcolor=gray!20,%
innertopmargin=0pt,%
innertopmargin=\topskip,%
splittopskip=\topskip,
ntheorem]{theorem}{Novelty Statement}
\usepackage{amssymb}
\usepackage{pifont}
\begin{document}

\preprint{AIP/123-QED}

\title[\em Zhai \& Yeo: Multiscale Fracture Mechanics of Graphene]{Multiscale mechanics of thermal gradient coupled graphene fracture: A molecular dynamics study}

\author{Hanfeng Zhai}
\affiliation{Sibley School of Mechanical and Aerospace Engineering\\ Cornell University}%
\author{Jingjie Yeo}

\email{\tt jingjieyeo@cornell.edu}

\affiliation{Sibley School of Mechanical and Aerospace Engineering\\ Cornell University}%

\date{\today}

\begin{abstract}
The thermo-mechanical coupling mechanism of graphene fracture under thermal gradients possesses rich applications whereas is hard to study due to its coupled non-equilibrium nature. We employ non-equilibrium molecular dynamics to study the fracture of graphene by applying a fixed strain rate under different thermal gradients by employing different potential fields. It is found that for AIREBO and AIREBO-M, the fracture stresses do not strictly follow the positive correlations with the initial crack length. Strain-hardening effects are observed for “REBO-based” potential models of small initial defects, which is interpreted as blunting effect observed for porous graphene. The temperature gradients are observed to not show clear relations with the fracture stresses and crack propagation dynamics. Quantized fracture mechanics verifies our molecular dynamics calculations. We provide a unique perspective that the transverse bond forces share the loading to account for the nonlinear increase of fracture stress with shorter crack length. Anomalous kinetic energy transportation along crack tips is observed for “REBO-based” potential models, which we attribute to the high interatomic attractions in the potential models. The fractures are honored to be more “brittle-liked” carried out using machine learning interatomic potential (MLIP), yet incapable of simulating post-fracture dynamical behaviors. The mechanical responses using MLIP are observed to be not related to temperature gradients. The temperature configuration of equilibration simulation employing the dropout uncertainty neural network potential with a dropout rate of 0.1 is reported to be the most accurate compared with the rest. This work is expected to inspire further investigation of non-equilibrium dynamics in graphene with practical applications in various engineering fields.
\end{abstract}

\maketitle
\begin{theorem}

\begin{itemize}\rm
    \item[]
    \item Using non-equilibrium molecular dynamics, graphene fracture is studied under temperature gradients with fixed strain rates to examine the effects of initial defect sizes, temperature differences, and interatomic forcefields.
    \item The stress-strain responses are highly dependent on the forcefield used, where the fracture stresses do not positively correlate with initial defect sizes for ``REBO-based" forcefields and strain-hardening effects are observed.
    \item The direction of fracture is not related to the temperature gradient.
    \item An abnormal form of fracture is observed for ``REBO-based" potentials, where the kinetic energy is transported along with the crack tips before fracture and is more frequently observed for longer initial pores of higher temperature differences.
    \item A comparative study of empirical MD forcefields with state-of-the-art {\em ab initio}-based machine learning potentials is also presented, where the limitations and the fracture characterizations are elaborated.
\end{itemize}
\end{theorem}

{\bf\textit{K}eywords }{Two-dimensional materials; nanomaterials; molecular dynamics; fracture; heat transfer; machine learning potentials}

\section{\label{intro}Introduction}

Two-dimensional materials are one of the fastest growing and active nanomaterials research areas, due to its exceptional mechanical \citep{jmr_graphene_mechanical, graphene_mechanical, nsr_review}, thermal \citep{thermal_book, 2d_thermal, aps_thermal}, electrical properties \citep{cornell_thesis_2008, electric_1}. Graphene is a 2D material with a single layer of carbon atoms arranged in a honeycomb lattice structure with $\rm sp^2$ bonds \citep{jj}. The successful synthesis of graphene \citep{first_graphene_paper} led to significant technological advances in graphene-based devices such as semiconductors \citep{semiconductor, semiconductor2}, batteries \citep{graphene_battery_1, graphene_battery_2}, biomedical devices \citep{biomed_1, biomed_2}, water desalination membranes \citep{water_1, water_2}, and many other industrial applications, largely because of its superior mechanical \citep{mech1, Thermal_Mechanical} and thermal properties \citep{thermal1, thermal2, thermal_3}.

More specifically, the high toughness \citep{huajian_review_fracture, Thermal_Mechanical, fracture_quantum}, strength \citep{strength1, strength2}, and thermal conductivity \citep{jj, jj_nanoscale} make graphene an ideal candidate for a broad variety of engineering applications. During the fracture of graphene \citep{huajian_review_fracture}, the nonlinear elastic regime plays a significant role in determining the strength of graphene \citep{elastic1, elastic2}. Both the strength and the presence of defects strongly influence graphene fracture \citep{nsr_review}. Topological defects like dislocations and grain boundaries can alter both the mechanical \citep{defect_mech1, defect_mech2, defect_mech3} and the thermal properties \citep{jj, nsr_review, defect_therm, defect_mechanical_thermal} of graphene. The effects on graphene fracture from the coupling of thermal and mechanical loads remain an interesting and ongoing research topic. Jangid and Kottantharayil \citep{heat_reconstruct} showed that methane gas treatment at a high temperature can reconstruct fractured graphene, considering one of the main reasons for graphene fracture is the electrical breakdown due to resistive heating. Liu et al. \citep{Liu_2020} tailored the microstructure of graphene composites to enable both high thermal conductivity and toughness. Most interestingly, Yo, Xu and Ding \citep{heat_treat} used both experimental  approaches and Monte Carlo simulations to show that multiple single-wall carbon nanotubes (SWCNT) under high-temperature heat treatment merged into new morphologies as temperature differences could break and reform carbon-carbon bonds.

The development of machine learning (ML) and data-driven methods enabled new advances in computational modeling and molecular simulations. One such example is the machine learning potential (MLP). Most MLPs adopt the pioneering concept by Behler and coworkers \citep{Behler_PRL, behler1, behler2} of utilizing neural networks to learn the molecular energy configuration based on first principle calculations to scale up \textit{ab initio} calculations. E and coworkers developed deep potential molecular dynamics (DeePMD) that employs the idea of \citep{Behler_PRL}'s formulation but further can train and infer atomic potential fields of different target materials implemented in state-of-the-art computational platforms \citep{deepmd, deepmd_prl}. Shapeev and coworkers developed machine learning interatomic potentials (MLIP) based on moment tensor operations  \citep{mlip}. Wen et al. used dropout matrices to thin the original neural networks for less uncertainty and named the method dropout uncertainty neural network (DUNN) \citep{dunn}. For the ease of fast implementation in PyTorch, Gao et al. developed a framework called TorchANI \citep{torchani}. Most recently, Jung et al. \citep{mlp_graphene_gangseob} developed an MLP using TorchANI specifically for graphene fracture. In training the MLPs, Jung et al. \citep{mlp_graphene_gangseob} and DUNN \citep{dunn, dunn_graphene} formulations include the differences of energy and forces in the loss function, whereas the MLIP \citep{mlip} and DeePMD \citep{deepmd} also include the (virial) stress. Arising from these important milestones, it is still unknown how these MLPs will perform when compared with empirical potentials in molecular simulations regarding the nonequilibrium fracture dynamics of graphene.

Inspired by the work of Yo, Xu, and Ding \citep{heat_treat} and Jangid and Kottantharayil \citep{heat_reconstruct}, an interesting question hence arises: how will thermal energy influence the process of graphene fracture? This question is significant in three aspects. (1) Theoretically, the fracture process under a thermal gradient is nonlinear and non-equilibrium in nature \citep{thermal_couple_fracture}, which is hard to either model or experiment with. Hence, describing the physical details is difficult, which will be elaborated on in the next paragraph. (2) Considering graphene's broad range of applications, graphene layer(s) subjected to thermal gradients is an omnipresent scenario, either as materials for batteries or semiconductors \citep{graphene_battery_1, graphene_battery_2, semiconductor, semiconductor2} and in which defects are largely unavoidable \citep{nsr_review, Araujo2012, defect_evidence}. Therefore, solving this problem has valuable industrial potential. (3) There are few related studies on this topic. A number of publications studied graphene's mechanical \citep{mechanical_temperature, mech1} or thermal properties \citep{thermal1, thermal2, thermal_3} either separately or measured related parameters under equilibrium state \citep{Thermal_Mechanical, mechanical_temperature, mechanical_thermal_monolayer}, but not the non-equilibrium thermo-mechanical coupling in graphene fracture. 

When a thermal gradient is induced in the graphene between the heat source and sink, the physical system is not in equilibrium \citep{thermal_couple_fracture}. At the molecular scale, the fracture of graphene is fundamentally the breaking of carbon-carbon bonds, which in essence is also a non-equilibrium process. To explore the mechanisms underlying such coupled processes, we use non-equilibrium molecular dynamics (NEMD) simulations to study the fracture behavior of a single graphene layer subjected to varying thermal gradients. We adopt four of the most commonly used empirical forcefields from current literature to model graphene: reactive bond order (REBO) \citep{rebo}, adaptive intermolecular REBO (AIREBO) \citep{airebo}, AIREBO-M \citep{airebo-m}, and optimized Tersoff \citep{opt-tersoff} forcefields. We study the graphene thermo-mechanical responses for fracture characterization. We also adopt the MLIP \citep{mlip} and DUNN \citep{dunn} to characterize the differences within MLPs and compare them with empirical potentials.

The manuscript is arranged as follows: in Section \ref{method} we briefly introduce the mathematical derivation of the empirical potentials (Sec. \ref{empirical_potentials}) and MLPs (Sec. \ref{ml_potentials}), as well as our numerical setup, including the problem formulation and the simulation details. The results are presented and discussed in Section \ref{result}, where the effects of the empirical potentials are elaborated in Sec. \ref{potential_effect}, the influence of thermo-mechanical coupling on the fracture process is proposed in Sec. \ref{result_couple}, the fracture dynamics are characterized in Sec. \ref{fracture_characterization} and the MLPs are compared in Sec. \ref{ml_potentials}. Finally, we make our conclusions in Section \ref{conclusion}.



\section{Methodology and modeling\label{method}}


\subsection{Empirical Interatomic Potentials\label{empirical_potentials}}


In molecular modeling of materials, interatomic potential energy functions, also known as forcefields or potentials, constitute the materials' overall physical properties. Empirical potentials, describing the atomic interactions based on symbolized empirical mathematical formulation, calculate the energy and potential spaces with interatomic motion based on Newtonian dynamics. Here, several widely applied empirical potentials are adopted for graphene in our modeling for comparison.

Generally, the atomic energies can be expanded in series as the sum of potentials, in which similar models can be viewed as an analog of Taylor series expansion. Based on these ideas, the energy of N interacting particles can be written as: \begin{equation}
    \begin{aligned}
    E = \sum_i V_i (\mathbf{r}_i) + \sum_{i<j} V_2 (\mathbf{r}_i, \mathbf{r}_j) + \sum_{i<j<k} V_3 (\mathbf{r}_i, \mathbf{r}_j, \mathbf{r}_k) + ...,
    \end{aligned}
\end{equation}where $\mathbf{r}_n$ is the position of the $n^{\rm th}$ particle, and $V_m$ is called the $m$-body potential, where $\sum_i V_i (\mathbf{r}_i)$ is the external potential. Detailed discussions can be found in Ref. \citep{Tersoff_1988}. Here, we briefly elaborate on the basic forms of interatomic potentials $V_{ij}$ and energies $E$ of different empirical models.

\subsubsection{Optimized Tersoff}

In the Tersoff proposition, \citep{tersoff1986, Tersoff_1988}, the potential was derived for covalently bonded structures fitted through parameterized rescaling \citep{ferrante1983, rose1983}. Such an interatomic potential has the form:\begin{equation}
    \begin{aligned}
    E^{\rm\textsc{Tersoff}} = \sum_i E^{\rm\textsc{Tersoff}}_i = \frac{1}{2} \sum_{i\neq j}V^{\rm\textsc{Tersoff}}_{ij},\\
    V^{\rm\textsc{Tersoff}}_{ij} = f^{\rm\textsc{Tersoff}}_C (r_{ij}) \left[ A e^{(- \lambda_i r_{ij})} - B_{ij} e^{(-\lambda_2 r_{ij})} \right]
    \end{aligned}\label{tersoff}
\end{equation}where $E$ is the total energy of the system, $E_i$ is site $i$'s site energy, to make the asymmetry of $V_{ij}$ more intuitive. $V_{ij}$ and $r_{ij}$ are the interaction energy and distance between atom $i$ and $j$ respectively. $A$, $B$, $\lambda_1$ and $\lambda_2$ are parameters with positive values, with $\lambda_1 > \lambda_2$. $f_c$ is the cutoff function to restrict potential ranges. The second term of $V_{ij}$ represents bonding, where $B_{ij}$ includes the bond order and hence depends upon the environment. The details of these parameters are provided in the Electronic Supplementary Information (ESI) and Ref. \citep{tersoff1986}.


Subsequently, Lindsay and Broido \citep{opt-tersoff} proposed an optimized version of the Tersoff potential, which captured  graphene's thermal properties more accurately compared with the original Tersoff, REBO, and AIREBO potentials \citep{potential_model}. The target parameters were optimized with chi-square minimization \citep{opt-tersoff, chi_opt}. The chi-square ($\chi^2$) is given by \begin{equation}
    \begin{aligned}
    \chi^2 = \sum_i \frac{\mathcal{U}_i - \mathcal{U}_{\mathbb B}}{\mathcal{U}^2_{\mathbb B}} \zeta_i  
    \end{aligned}
\end{equation}where $\mathcal{U}_{\mathbb B}$ are benchmark parameters used in the fitting process, which can be based on first-principal calculations \citep{chi_opt} and/or experiments \citep{opt-tersoff}. $\mathcal{U}_i$ are the corresponding values obtained from the original Tersoff potential, and $\zeta_i$ are weighting factors that determine the relative importance of $\mathcal{U}_i$ in the fitting process. In our approach, the fitted parameters for the optimized Tersoff potential are given in the ESI.

\subsubsection{Reactive Bond Order (REBO)}
The reactive bond order (REBO) potential was first proposed by Brenner \citep{Brenner1990, Brenner1992}, which is an exclusively short-ranged potential \citep{airebo}. The interaction of two atoms is computed only when their distance is less than a covalent-bonding cutoff $r_{ij}^{\rm max}$, where the interaction follows: \begin{equation}
    E_{ij}^{\rm REBO} = \sum_{j \neq i} f^{\rm REBO}_c(r_{ij}) \left[V_R(r_{ij}) + \Bar{b}_{ij} V_{ij}^A\right]\label{rebo_eq}
\end{equation}where $V_{ij}^R$ and $V_{ij}^A$ are the repulsive and attractive pairwise potentials between atoms $i$ and $j$ determined from their interatomic distance, $r_{ij}$. $\bar{b}_{ij}$ is the many-body term (See Section 1.2 in ESI). The repulsive $V^{\rm R}$ and attractive $V^{\rm A}$ terms take the form \citep{Brenner1990}:\begin{equation}
    \begin{aligned}
    V_{ij}^{\rm R} = \sum_{n = 1}^3 B_n e^{\beta_n r}\\
    V^{\rm A}_{ij} = \left( 1 + \frac{Q}{r}\right)A e^{\alpha r}
    \label{attract}
    \end{aligned}
\end{equation}Full information on the pertinent parameters is provided in the ESI. Note that $V^{\rm A}_{ij}$ is switched off for long-ranged atomic interactions through bond weights. More details can also be found in Refs \citep{airebo, Brenner1992, Brenner1990}.



\subsubsection{Adaptive Intermolecular REBO (AIREBO)}

While successful in describing intramolecular interactions, the REBO potential still lacks the inclusion of intermolecular interactions. Stuart et al. \citep{airebo} further proposed the adaptive intermolecular REBO (AIREBO) method, adding Leonard-Jones (LJ) and torsional interactions to the total potential:\begin{equation}
    E^{\rm AIREBO} = \frac{1}{2} \sum_i \sum_{j \neq i} \left[ E^{\rm REBO}_{ij} + E^{\rm LJ}_{ij}+ \sum_{k \neq i,j} \sum_{l \neq i,j,k} E_{kijl}^{\textsc{Torsion}}\right]\label{airebo_eq}
\end{equation}where the detailed forms of $E_{ij}^{\rm LJ}$ and $E_{kijl}^{\rm torsion}$ and the corresponding $V^{\rm LJ}_{ij}(r_{ij})$ and $V_{kijl}^{\textsc{Torsion}}(r_{ij})$ are given in the ESI. The detailed derivation of these equations can be found in Ref. \citep{airebo}.

\subsubsection{AIREBO-M}

Even with the added intermolecular terms, the AIREBO potential was still unable to accurately model high-pressure systems due to extremely strong repulsive forces under such conditions. O'Connor, Andzelm, and Robbins \citep{airebo-m} replaced the LJ interactions with the Morse potential to more accurately describe the intermolecular interactions:\begin{equation}
    V^{\textsc{Morse}}_{ij}(r) = -\epsilon_{ij} \left[ 1 - \left(1 - e^{\alpha_{ij} \left( r -r^{eq}_{ij}\right)}\right)^2\right] 
\end{equation}where the depth and location of the minimum energy are defined through $\epsilon$ and $r^{eq}$. $\alpha$ modifies the curvature of the potential energy. 

The total energy and the potential energy can then be obtained by solving the Schrödinger Equation \citep{morse}\begin{equation}
    - \frac{\hbar}{2m} \frac{d^2\psi}{dx^2} + V^\textsc{Morse} \psi= E^{\textsc{Morse}} \psi
\end{equation}where $\psi$ is the wave function, $\hbar$ is the Planck constant, and $m$ is the particle's mass.

Hence, the final form of the total energy of AIREBO-M potential is:\begin{equation}
    E^{\rm AIREBO-M} = E^{\rm REBO} + E^{\textsc{Morse}} + E^{\textsc{Torsion}}\label{airebo-m_eq}
\end{equation}where detailed parameterization and definition of the Morse potential can be found in the ESI and Ref. \citep{airebo-m}.

The four empirical potentials introduced herein were employed in modeling nanoporous graphene fracture under thermal gradients with high strain rate loading for comparison and unveiling the underlying mechanism and the physics.

\subsection{Machine Learning $\bm{ab\ initio}$ Potentials\label{ml_potentials}}

The core idea of MLPs is to employ ML (neural networks in our cases) as an approximator to scale up molecular interactions based on quantum-mechanical calculations. Most state-of-the-art MLP models follow the pioneering work conducted by Behler and coworkers \citep{Behler_PRL, behler1, behler2}, which construct the {\em ab initio} computational domain using the atomic configurations as input for the ML model to construct the surrogates with energy fields as output. The general supervised learning task is formulated and the approximator (i.e., neural networks, Gaussian process) is trained on data based on density functional theory (DFT), {\em ab initio} molecular dynamics (AIMD), or other first principle methods. The learned energy fields can then be extended to calculate the interactions at the molecular level based on Newtonian dynamics. Here, two widely used MLPs, MLIP \citep{mlip} and DUNN \citep{dunn}, are adopted to benchmark the calculation of graphene properties.

\subsubsection{Machine-Learning Interatomic Potentials (MLIP)}

The MLIP model was first proposed by Shapeev and coworkers \citep{mlip, mlip_mag_vibration} and later implemented in graphene \citep{carbon_ml, advmat}. They apply 
moment tensor potentials (MTP) to seamlessly accelerate first principle calculations and incorporate active learning strategies for more efficient training and model construction. The total energy takes the form, 
\begin{equation}
    E^{\rm MTP} = \sum_{i=1}^n V_i(\mathbf{r}_i)\ \longrightarrow\ V_i(\mathbf{r}_i) = \sum_\alpha \xi_\alpha B_\alpha (\mathbf{r}_i)
\end{equation} Here, the function $V$ is linearly expanded through a set of basis function $B_\alpha$. $\xi = \{\xi_\alpha\}$ are parameters obtained through fitting to the training sets. Shapeev and coworkers then introduce moment tensor descriptors and construct the basis functions from the level of these moments. More details can be found in the ESI and their tutorial paper \citep{mlip}. 

Suppose the quantum-mechanical energy $E^{\rm QM}$ are known for training, with their corresponding stress tensors $\sigma^{\rm QM}$. The whole learning process can be viewed as using a neural network (NN) as an approximator to fit the known energy and stresses. If we denote the fitting parameters as $\theta$, the fitting procedure can be simplified as:\begin{equation}
   \mathcal{L} = \sum_{k=1}^K \left[ w_e \mathcal{E}(E^{\rm MTP}, E^{\rm QM}) + w_f\sum_{i=1}^{N_k} \mathcal{E}\left(\mathbf{f}^{\rm MTP}_i, \mathbf{f}^{\rm QM}_i\right) + w_s \mathcal{E}\left(\sigma^{\rm MTP}, \sigma^{\rm QM}\right) \right] \xrightarrow{} \min_\theta 
\end{equation}where $N_k$ is the atomic numbers in the $k^{th}$ configuration. $w_e$, $w_f$, and $w_s$ are weights for energies, forces, and stresses. $\mathcal{E}$ is the error measurements, i.e., mean-square errors (MSE), root MSE, and absolute errors. $\mathcal{L}$ is the loss function to be minimized during optimization. $\theta$ are the hyperparameters for the MTP predicted physical properties evolving during optimization. The detailed forms are given in the ESI and further derivation in Ref. \citep{mlip}. Here, MLIP is used to replace traditional empirical potentials to benchmark the simulations of porous graphene fracture.

\subsubsection{Dropout Uncertainty Neural Network (DUNN)}

Even though MLP methods are useful for scaling up molecular simulations with {\em ab initio} accuracy, they lack interpretable models for the observed phenomena, leading to unknown accuracy outside the training set. Wen and Tadmor dealt with these limitations by proposing the DUNN model by eliciting a dropout matrix \textbf{D} to the original energy formulation by Behler and coworkers \citep{Behler_PRL, behler1, behler2},
\begin{equation}
    \begin{aligned}
    E^{\rm DUNN} = \sum_{i=1}^N E^{\rm DUNN}_i,\\ \longrightarrow\ E^{\rm DUNN}_i = h\left[h[\mathbf{y}_0(\mathbf{D}_1\mathbf{W}_1) + \mathbf{b}_1](\mathbf{D}_2\mathbf{W}_2) + \mathbf{b}_2\right](\mathbf{D}_3\mathbf{W}_3) + \mathbf{b}_3
    \end{aligned}\label{dunneq}
\end{equation}where $\mathbf{D}_i$ is a square diagonal binary matrix of integers 0 or 1, in which the diagonal follows the Bernoulli distribution \citep{dunn}. With the formulation by Behler and coworkers \citep{behler2}, the input layer $\mathbf{y}_0$ (where $y_0^j$ denotes the $j^{\rm th}$ neurons in the layer) is transformed through $N_{\rm desc}$ descriptors $g^j (\mathbf{r}_i)$ satisfying the symmetry requirements, with a specified cutoff radius. Note that $\mathbf{r}_i$ can be viewed as describing the local environment within the cutoff, such that  \begin{equation}
    y_0^j = g^j (\mathbf{r}_i),\ j = 1,2,...,N_{\rm desc}
\end{equation}

Now, one can define Equation (\ref{dunneq}) as the dropout NN with the updated weights $\Tilde{\mathbf{W}}_i = \mathbf{D}_i\mathbf{W}_i$, for which the new model is interpreted by Wen and Tadmor as a Bayesian model. 
In such a model, the prior distribution $p(\omega)$ over parameters $\omega = \{\Tilde{\mathbf{W}}_1, \Tilde{\mathbf{W}}_2, \Tilde{\mathbf{W}}_3, \mathbf{b}_1, \mathbf{b}_2, \mathbf{b}_3 \}$ induces the predictive distribution concerning the likelihood $p(\mathcal{Y}\ |\ \mathcal{X}, \omega)$ for training data of $(\mathcal{X}, \mathcal{Y})$.
\begin{equation}
    \begin{aligned}
    p(\omega\ |\ \mathcal{X}, \mathcal{Y}) \propto p (\mathcal{Y}\ |\ \mathcal{X}, \omega)p(\omega), \\ \longrightarrow p (\mathbf{z}\ |\ \mathbf{x}^*, \mathcal{X}, \mathcal{Y}) = \int p(\mathbf{z}\ |\ \mathbf{x}^*, \omega) p(\omega\ |\ \mathcal{X}, \mathcal{Y}) d\omega
    \end{aligned}
\end{equation}where $\bf z$ is the quantity of interest, and $\bf x^*$ are the descriptors for a configuration associated with \textbf{z}. Note that the training process of DUNN differed from MLIP by the fact that the loss function only contains the differences of energy and forces with no stress included \citep{dunn}. 

Here, three different types of DUNN \citep{dunn1, dunn2, dunn3} with different dropout rates were adopted to study the temperature distribution at equilibrium using a small portion of the entire graphene sheet. The details are elaborated further in Section \ref{simulation_setup} and the ESI.



\subsection{Simulation Setup\label{simulation_setup}}

\begin{figure}[htbp]
    \centering
    \includegraphics[scale=0.23]{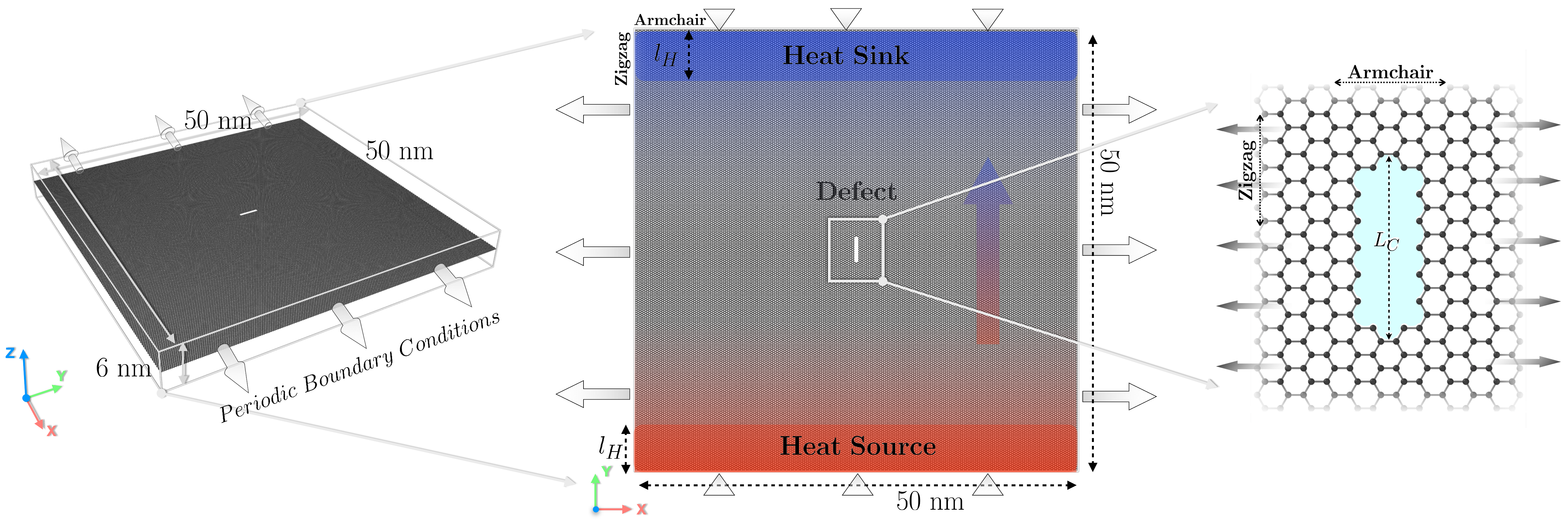}
    \caption{The schematic of the simulation setup. A single graphene layer with a defect is constrained in the simulation box, with periodic boundary conditions applied on each side. A temperature gradient is enforced in the Y direction (green arrow) by fixing two constant temperature regions (heat sink and heat source) of height $l_H$. A tensile strain rate of $10^{10} \ \rm s^{-1}$ is applied in the X direction (red arrow). The size of the simulation box is $\rm 50 \times 50 \times 6\ nm$ ($\mathsf{X\times Y \times Z}$). The defect is a symmetric atomic vacancy with a width of 0.71 nm. $L_C$ denotes the length of the crack (or defect) and the defect is generated by continuously creating a double vacancy and removing their adjacent carbon atoms.}
    \label{schematic}
\end{figure}

The MD model consists of a three-dimensional simulation box with X and Y dimensions of 50 nm and the Z dimension (height) of 6 nm with full periodic boundary conditions (Figure \ref{schematic}). The X direction is the armchair direction and {Y} is the zigzag direction. A thermal gradient is enforced in the Y direction using a heat source and sink placed at the lower and upper portions respectively. At the center of the graphene layer, a defect of different lengths is introduced to account for possible sizing effects from this pre-crack. To propagate the crack, a strain rate of $10^{9} \ \rm s^{-1}$ is applied in the X direction, as indicated by the gray shaded arrows in Figure \ref{schematic}. 

To investigate the thermo-mechanical coupling behavior, two parameters are tuned in the simulation: the precrack length $L_C$, and the temperature differences between the heat source and sink $\rm\Delta T$. Five different precrack lengths (1.7217, 3.1974, 4.1812, 5.6569, and 8.1164 nm) with a width of 0.71 nm, and four different thermal gradient values (0 K, 100 K, 200 K, and 300 K) are probed in the simulations.

The mechanical properties are characterized by stress-strain responses. Using the deformation gradient tensor $\mathbf{F} = \frac{\partial \mathbf{x}}{\partial\mathbf{X}}$ described by the reference and current configurations $\mathbf{X} = (X_1, X_2, X_3)$ (can be also written as $\mathbf{X} = (\rm X, Y, Z)$, here use $X_i$ for the ease of notations for strain representations) and $\mathbf{x} = (x_1, x_2, x_3)$ (can be also written as $\mathbf{x} = (\rm x, y, z)$), the constitutive model can be written as $\sigma = \Phi \left(\mathbf{F}(\mathbf{X}, t), \mathbf{X}\right)$. From \textbf{F} one can derive the displacements, $\mathbf{u} = \mathbf{x} - \mathbf{X}$, from which one can obtain the strain in 3D with indicial notation: $\epsilon_{ij} = \frac{1}{2}\left(\frac{\partial u_i}{\partial X_j} + \frac{\partial u_j}{\partial X_j}\right)$. The strain rate will then be

\begin{equation}
    {\dot {\epsilon }}(t)={\frac {d\epsilon }{dt}}={\frac {d}{dt}}\left({\frac {x_i(t)-X_i}{X_i}}\right)
\end{equation} By applying a constant strain rate, the corresponding stress-strain response of the graphene layer can be determined, where the yield stress is $\sigma_Y = \mathtt{max}(\sigma(t))$, and the corresponding yield strain takes the form $\epsilon_Y = \Phi^{-1}(\sigma_Y)$.

By performing the simulations using the empirical potentials, three properties are of core interest when studying the mechanism of such non-equilibrium fracture dynamics: the separate effects of the pre-crack lengths and thermal gradients, and the coupled effects of the thermo-mechanical mechanisms on the fracture of graphene. Here, we apply a high strain rate $\dot{\epsilon} = 10^9 {\ \rm s^{-1}}$ ($10^{-3} {\ \rm ps^{-1}}$) according to the work of Zhao and Aluru \citep{jap}, as we hope to (1) benchmark our mechanical responses and compare the results; (2) investigate the coupling mechanisms during the fracture process under this non-equilibrium condition. Our main goal is to explore the mechanical responses while considering (1) exploring the variations between potential models, and (2) unraveling the molecular physical details independent of the errors induced by different computational modeling methods, i.e., the interatomic potential employed.

\begin{figure}[htbp]
    \centering
    \includegraphics[scale=0.25]{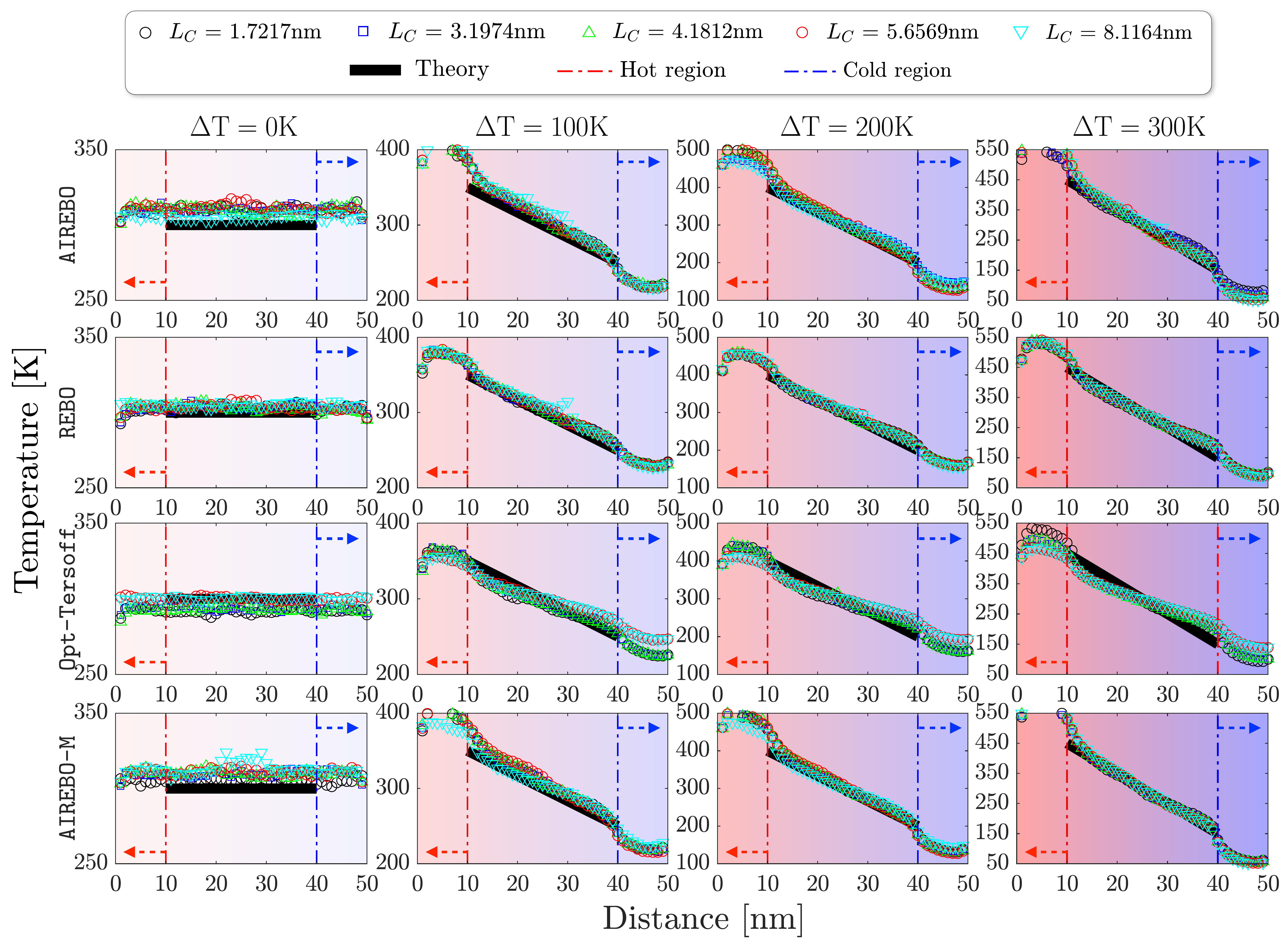}
    \caption{The temperature distribution within the single graphene layer after equilibrating for 10,000 steps. The scattered dots in different colors represent different graphene layers with varying pre-crack lengths, as indicated in the legend. The red dashed line and arrow denote the boundary of the ``heat source" region with a higher temperature, while the blue dashed line and arrow denote the boundary of the ``heat sink" region with a lower temperature. The linear fit to the temperature distribution is marked as a black solid line. The four columns indicate four different temperature gradients and the four rows are the temperature distribution under different potential fields.}
    \label{temperature_grad}
\end{figure}

After running for 10,000 steps, the temperature distribution along the Y position is shown in Figure \ref{temperature_grad}. The four subfigures indicate the temperature distribution along with the {Y} direction position. We conclude that the temperature gradient is linear within the mid-region where the crack will propagate. Following the equilibration, 500,000 steps of tensile loading were carried out under the constant thermal gradient. The simulation was carried out using the NVE ensemble. (See Section 7 in ESI for the details of implementation in LAMMPS)

For benchmarking the MLPs, two simulation cases were set: (1) when benchmarking the MLIP potential model, we directly replace the empirical potentials with MLIP and carried out the same simulations (See Section 2.2 in ESI for technical implementation details). (2) In our attempts the DUNN model could not handle deforming boxes and high-temperature gradients: the simulation breaks into errors from the DUNN potentials; and also tends to be more computationally consuming, we hence create a smaller simulation box (length $\sim \frac{1}{5}$ of the original length) and only ran the 10,000 steps of equilibration with zero temperature gradient following the same procedure as before to test the model's ability to recreate the thermal conditions. Four MLP models, i.e., MLIP \citep{mlip}, DUNN v1 \citep{dunn1}, DUNN v2 \citep{dunn2}, DUNN v3 \citep{dunn3}, are all employed in this case. Note that DUNN v1, v2, and v3, stand for the DUNN with different dropout ratios of 0.1, 0.2, and 0.3, respectively. The details can be found in Ref. \citep{dunn}.

\section{Results and discussion\label{result}}

\subsection{Influence of Interatomic Potentials on Thermo-mechanical Responses \label{potential_effect}}

\begin{figure}
    \centering
    \includegraphics[scale=0.27]{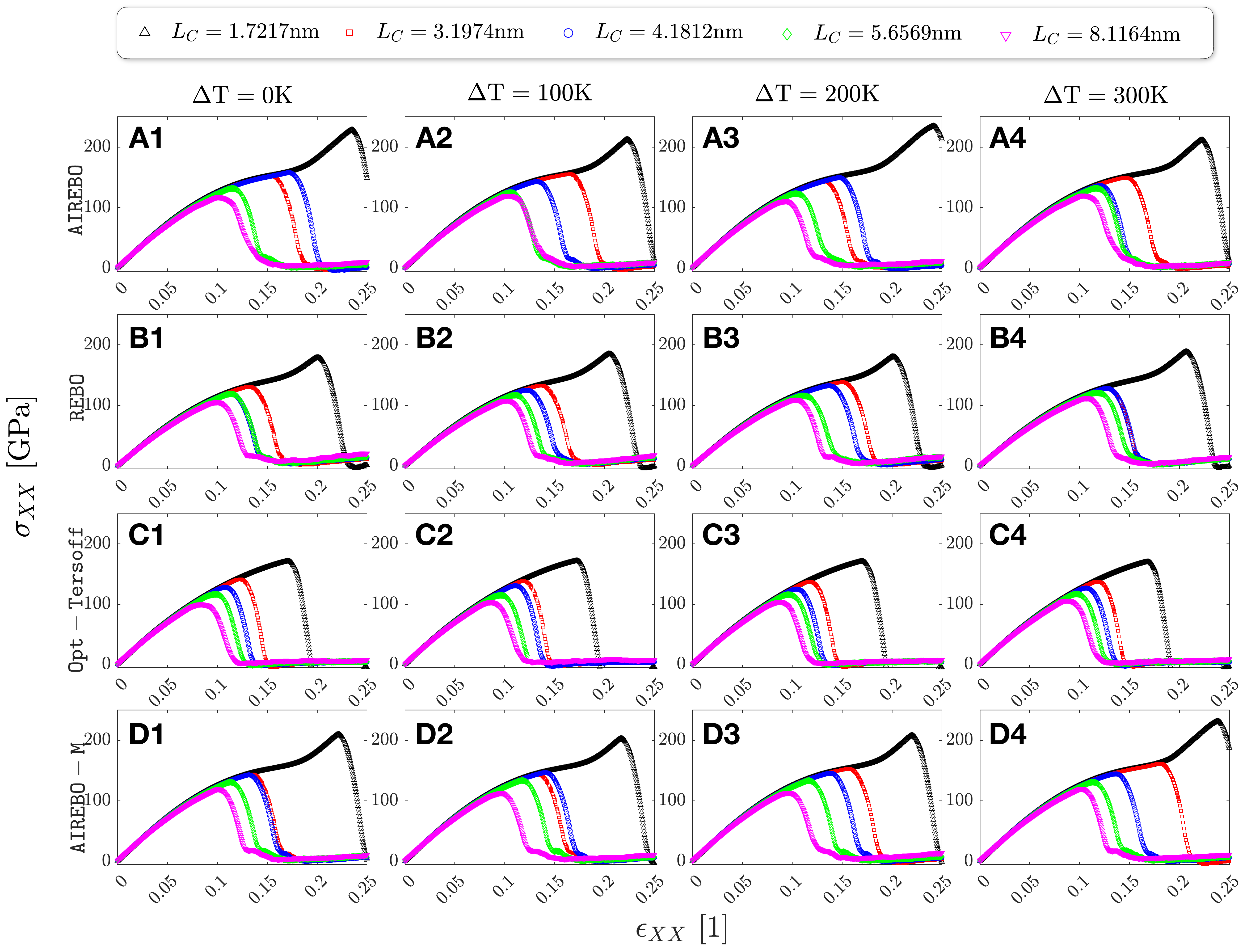}
    \caption{The stress-strain responses of the graphene sheet during non-equilibrium, high strain rate tensile tests under thermal gradients. Subfigures {\bf A, B, C, D} denote the simulations using AIREBO, REBO, optimized Tersoff, and AIREBO-M potential fields, respectively, as indicated in the labels. The corresponding suffixes {\bf 1, 2, 3, 4} denote different temperature gradients as indicated in the titles. Lines of different colors denote graphene sheets of different initial defects as shown in the legend.}
    \label{stress_strain}
\end{figure}

Figure \ref{stress_strain} shows the mechanical responses of graphene sheets while varying the initial defect lengths, temperature gradients, and interatomic potential fields. Intuitively, longer initial defects should result in graphene fracturing at lower stresses. But by observing Figure \ref{stress_strain} {\bf A} \& \textbf{D}, such a trend is not strictly obeyed: the blue and red dots shift back and force at different temperature gradients. In contrast, from both Figure \ref{stress_strain} \textbf{B} and \ref{stress_strain} \textbf{C}, the pre-crack lengths correlate with fracture stresses. We deduce that ``REBO-based" potentials, i.e., REBO, AIREBO, AIREBO-M, exhibit non-intuitive results: the fracture stresses are not strictly positively correlated with initial crack length. This point will be discussed further in our characterization of the fracture profile in Section \ref{fracture_characterization}. Another interesting phenomenon is that the simulations employing the ``REBO-based" potentials display strain-hardening effects for graphene with small initial defects, shown by the black dots in Figure \ref{stress_strain} \textbf{A}, \textbf{B} and \textbf{D}. We proffer two explanations: (1) The coupling effect of relatively high strain rate and strong attraction between atoms. From Equation (\ref{attract}), we know the attractive forces are switched off for long distances in the REBO potential. Under certain strain rate loading, at a specific strain when the interatomic distance is still within the cutoff range, the interatomic attraction that still resists the applied loading contributes to the stress increase as the strain-hardening effect we observed. (2) The transverse bond energy in the X direction further resists the loading. The strain hardening effect is only observed when the initial defect is small in our simulations, where chemical bonds in the X direction help resist the loading. This point will be elaborated further in Section \ref{result_couple}. Such strain-hardening phenomena are also observed in MD simulations of nanoporous graphene \citep{grossman, iop_conference}, graphene nanoribbons \citep{phys_lett_a}, multilayer graphene \citep{jap_strain_hardening}, which agree with our findings here.

By comparing Figure \ref{stress_strain} \textbf{A} to \textbf{D}, the temperature did not affect the mechanical responses for different potentials. By defining the fracture stress to be the highest value during loading and the corresponding strain to be the fracture strain, in Figure \ref{stress_strain} \textbf{C}, higher temperature gradients reduce the differences between fracture stresses and strains with different initial defects if the optimized Tersoff potential is used. In Figures \ref{stress_strain} \textbf{A} \& \textbf{B}, the AIREBO and REBO potentials approximate such a trend as well. In Figure \ref{stress_strain} \textbf{D} using the AIREBO-M potential, higher temperature gradients seem to increase the differences between fracture stresses and strains of different initial defect sizes. In short, no clear mechanism can be observed to relate fracture stresses and strains to temperature across the different potentials. The AIREBO and AIREBO-M potentials are reported to exhibit higher fracture stress values and corresponding strains, with the optimized Tersoff and REBO potentials displaying lower fracture stresses and strains.

\subsection{Thermo-Mechanical Coupling Mechanism\label{result_couple}}

\begin{figure}[htbp]
    \centering
    \includegraphics[scale=0.3]{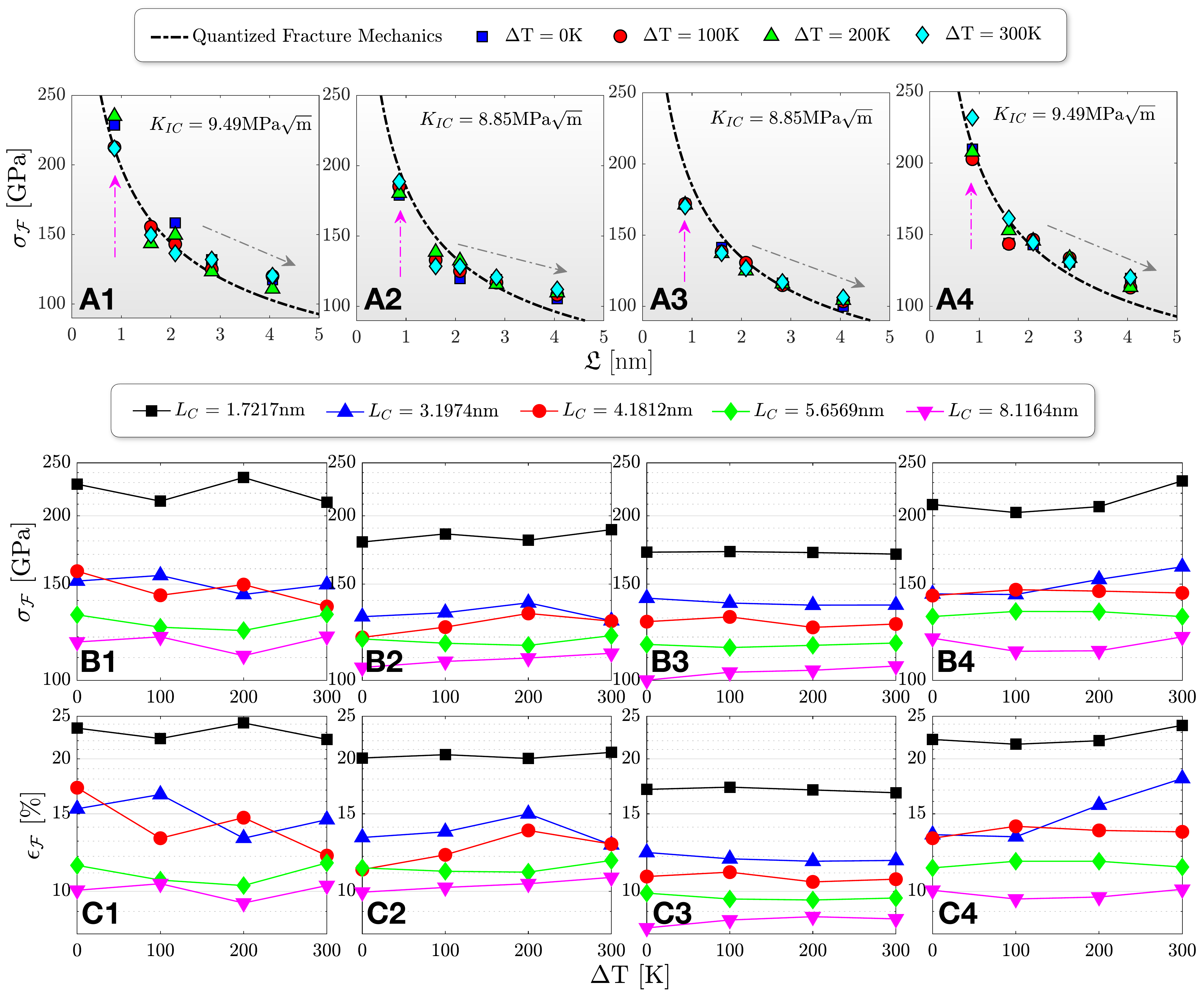}
    \caption{The fracture stresses and strains ($\sigma_\mathcal{F}$ \& $\epsilon_\mathcal{F}$) of the graphene layer with varying initial defect lengths and temperature differences. \textbf{A}. The fracture stresses in relation to initial defect length, with the fitted curve by QFM marked in a black dashed line. The corresponding values under different temperature gradients are marked in different shaped and colored markers. The fitted fracture intensities $K_{IC}$ values are provided on the right-top corners of each subfigure. \textbf{B}. The relations between fracture stress and temperature gradients. \textbf{C}. The relations between fracture strains and temperature gradients. Note that for \textbf{B} and \textbf{C}, each colored dotted line indicates a specific initial crack length, $L_C$, provided in the legend above.}
    \label{linear_fit}
\end{figure}

Figure \ref{linear_fit} shows the relation between the fracture stresses and strains to different initial defect sizes and temperature gradients (See Figure S5 in Section 6 in ESI for 3D data visualization contour plot). To validate our results that relate fracture stresses to defect sizes (in Figure \ref{linear_fit} \textbf{A}), we apply quantized fracture mechanics (QFM) to determine the fracture intensity $K_{IC}$ for comparison with literature values. In the work by Pugno and Ruoff \citep{qfm}, to describe discrete crack propagation, Griffith's criterion can be quantized using QFM: an energy-based method that could accurately describe fracture observed in molecular dynamics (MD) simulations \citep{qfm_md}. If we denote the smallest length of pre-crack that will propagation to be $L_0$, for graphene along the zigzag direction, $L_0 = 0.246$ nm; the initial crack (defect region of graphene) has a length of $L_C = 2\mathfrak{L}$, where $\mathfrak{L}$ is the half of the pre-crack length, used in the derivation of QFM theory for easier benchmarking; and $\rho$ is the tip radius, which in our case $\rho = 0.265$ nm (See Figure S2 of Section 3.2 in ESI and Ref. \citep{jap}). In continuum-based linear elastic fracture mechanics (LEFM), fracture occurs when the stress intensity equals its critical value, $K^{\rm LEFM}_I = K_{IC}$ \citep{fracture_criterion}. In QFM, the crack propagates when \citep{qfm}\begin{equation}
    K^{\rm QFM}_I = \sqrt{\frac{1}{L_0}\int_\mathfrak{L}^{\mathfrak{L}+L_0} \left[{K^{\rm LEFM}_I} (\mathfrak{L}) \right]^2d \mathfrak{L}} = K_{IC}
\end{equation}Substituting this $K_{IC}$, the fracture stress in QFM writes $\sigma_\mathcal{F}(\mathfrak{L}) = \frac{K_{IC}}{\sqrt{\pi (\mathfrak{L} + L_0/2)}}$, By extending this fracture stress from sharp to blunt cracks \citep{frac_1_deriv, frac_2_deriv}, an asymptotic correction for small tip radii is needed in the form \citep{qfm} :

\begin{equation}
    \sigma_\mathcal{F}(\mathfrak{L}, \rho) = K_{IC} \sqrt{\frac{1 + \frac{\rho}{2L_0}}{\pi (\mathfrak{L} + \frac{L_0}{2})}}\label{qfm_eq}
\end{equation}

We use Equation (\ref{qfm_eq}) to calculate the critical stress intensity. We find that the $K_{IC}$ values for the four potentials are approximately 9.49, 8.85, 8.85, 9.49$\ \rm MPa\sqrt{m}$, similar to the value of 9.2$\pm 0.8\ \rm MPa\sqrt{m}$ of fracture direction along the armchair direction and $\sim73^{\circ}$ relative to the loading direction of the monolayer in Ref. \citep{adv_mat_KIC} and $10.7 \sim 14\ \rm MPa\sqrt{m}$ in CVD-grown graphene \citep{fracture_toughness_10}. Yet, these results are larger than the experimental value of 4.0$\pm 0.6\ \rm MPa\sqrt{m}$ \citep{nat_comm_KIC_smaller_value} and smaller than the value of $12\pm 3.9\ \rm MPa\sqrt{m}$ in multilayer graphene \citep{fracture_toughness_12}. By benchmarking against values by Zhao \textit{et al} \citep{adv_mat_KIC}, our calculated values have relative errors of 3.15\%, 3.80\%, 3.80\%, and 3.15\%, respectively, indicating that our calculations are generally accurate. Based on our fitted $K_{IC}$ using QFM (Figure \ref{linear_fit} \textbf{A}), we find an excellent match with our simulation data. Furthermore, to verify the fitting accuracy in Figure \ref{linear_fit}, we compute the $R^2$ scores (a.k.a. coefficient of determination) w.r.t. each potential model and temperature differences using the following equation and generate Table \ref{r2score_qfm}:\begin{equation}
    R^2 = 1- \frac{\sum\left(\sigma_{\mathcal{F}}^{\rm MD} - \sigma_{\mathcal{F}}^{\rm QFM} \right)^2}{\sum\left({\tt mean}\left(\sigma_{\mathcal{F}}^{\rm MD}\right) - \sigma_{\mathcal{F}}^{\rm QFM}\right)^2}
\end{equation}where $\sigma_{\mathcal{F}}^{\rm MD}$ and $\sigma_{\mathcal{F}}^{\rm QFM}$ denote the fracture stresses computed from MD (using different empirical potential models) and QFM, respectively; and $\tt mean(\cdot)$ denote the mean values of the total samples computed. Table \ref{r2score_qfm} shows that the fitted QFM curves are generally accurate for the MD simulation data, cross-verified both the accuracy of QFM and our MD simulations.

\begin{table}[htbp]
    \centering
    \begin{tabular}{c|c c c c}\hline
         & $\rm\Delta T = 0$ K & $\rm\Delta T = 100$ K & $\rm\Delta T = 200$ K & $\rm\Delta T = 300$ K\\\hline
        \tt AIREBO & 0.8741 & 0.9538 & 0.8740 & 0.9249 \\
        \tt REBO & 0.8395 & 0.8938 & 0.8964 & 0.8623 \\
        \tt Opt-Tersoff & 0.8841 & 0.8676 & 0.8780 & 0.8495\\
        \tt AIREBO-M & 0.9117 & 0.9101 & 0.9101 & 0.8967 \\\hline
    \end{tabular}
    \caption{The $R^2$ scores calculated for estimating the fitting accuracy of QFM w.r.t. MD simulations using different empirical potentials under different temperature differences.}
    \label{r2score_qfm}
\end{table}

\begin{figure}[htbp]
    \centering
    \includegraphics[scale=0.22]{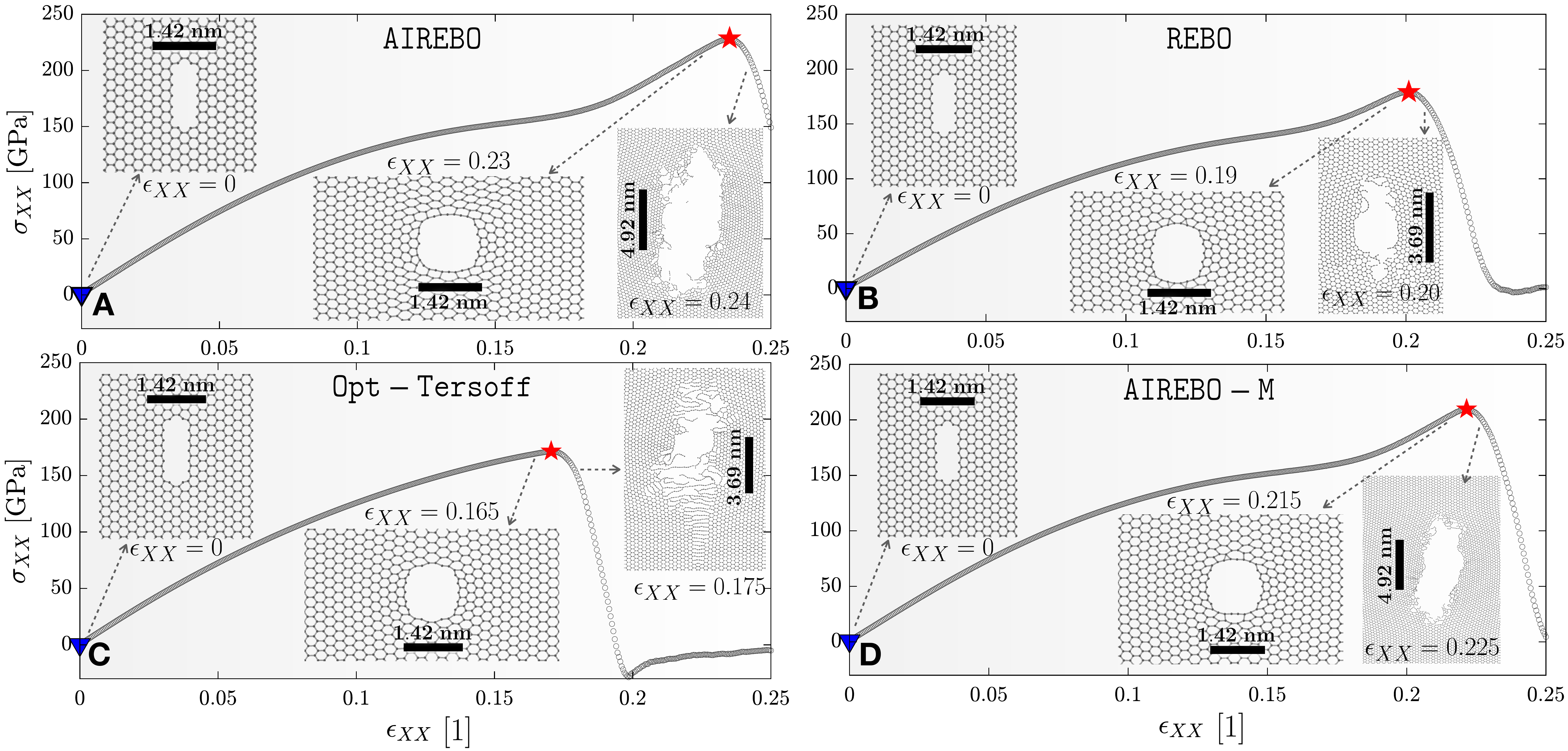}
    \caption{The stress-strain responses of the graphene sheet for a small initial defect ($L_C = 1.7217$ nm) using the four different potential fields. The four subfigures {\bf A, B, C, D} illustrate the deformation and fracture profile along the defective area marked with the corresponding strain values. The red star denotes the point where the fracture occurs. The inset at the blue triangular dot is the initial morphology of the defect.}
    \label{explain_small_pore}
\end{figure}

We also provide a new perspective regarding the phenomenon observed in Figure \ref{linear_fit} \textbf{A}: as the initial crack lengths get larger, the fracture stresses decrease more mildly. To illustrate, the gray dashed arrows highlight the milder decreases as the crack length increases, whilst the pink dashed arrows highlight the nonlinear, steeply decreasing trends in shorter cracks. We proffer that for shorter cracks, the bonds adjacent to the crack tip can withstand a significant amount of the horizontal (X direction) component of the force causing crack opening at the crack tip. Hence, the fracture stress increases and lead to crack rounding as well as the nonlinear increase of the fracture stress when the pre-crack length decreases. For longer initial cracks, forces are mostly concentrated at the crack tip itself and the adjacent bonds share less of the applied loads, leading to milder fracture stresses when the initial cracks are longer. To illustrate, Figure \ref{explain_small_pore} shows the fracture process with initial crack $L_C = 1.7217$ nm corresponding to the stress-strain responses. Comparing the morphology of the cracks for all four potentials before and after the fractured moment (red star), it can be observed that the overall crack widths are larger as the bonds adjacent to the crack tip share the loads, leading to rounding of the cracks. Moreover, by comparing Figure \ref{explain_small_pore} \textbf{A} \& \textbf{D} with \textbf{B} \& \textbf{C}, we observe that, for AIREBO and AIREBO-M, the fracture begins at a larger strain compared with REBO and optimized Tersoff: (1) the longest pulled widths are longer by observing the defect morphology preceding fracture and (2) the fracture strains denoted by the red star are higher.

From Equations (\ref{tersoff}) and (\ref{rebo_eq}), the REBO and Tersoff models follow similar formulations: the atomic interactions are linear combinations of repulsive and attractive interactions. The attractive terms in REBO (Equation (\ref{attract})) are switched off for long-distance interactions and the weight function $f^C_{ij}$ ensures interatomic interactions are zero outside the cutoff range. From Equations (\ref{airebo_eq}) and (\ref{airebo-m_eq}), the AIREBO and AIREBO-M potentials contain additional terms for the torsion energy, as well as the Leonard-Jones and/or Morse energies, in comparison with the REBO model. Hence, we propose that these extra energetic terms in the AIREBO and AIREBO-M potentials can model the dynamic loading conditions more accurately compared to the optimized Tersoff and REBO models in Figure \ref{explain_small_pore}. Figures \ref{explain_small_pore} {\bf B} \& \textbf{C} indicate that no clear relationship can be inferred between the temperature gradients and the fracture stresses and strains, thus agreeing with the trend we observed in Figure \ref{stress_strain}.



\subsection{Fracture Characterizations\label{fracture_characterization}}

\begin{figure}
    \centering
    \includegraphics[scale=0.23]{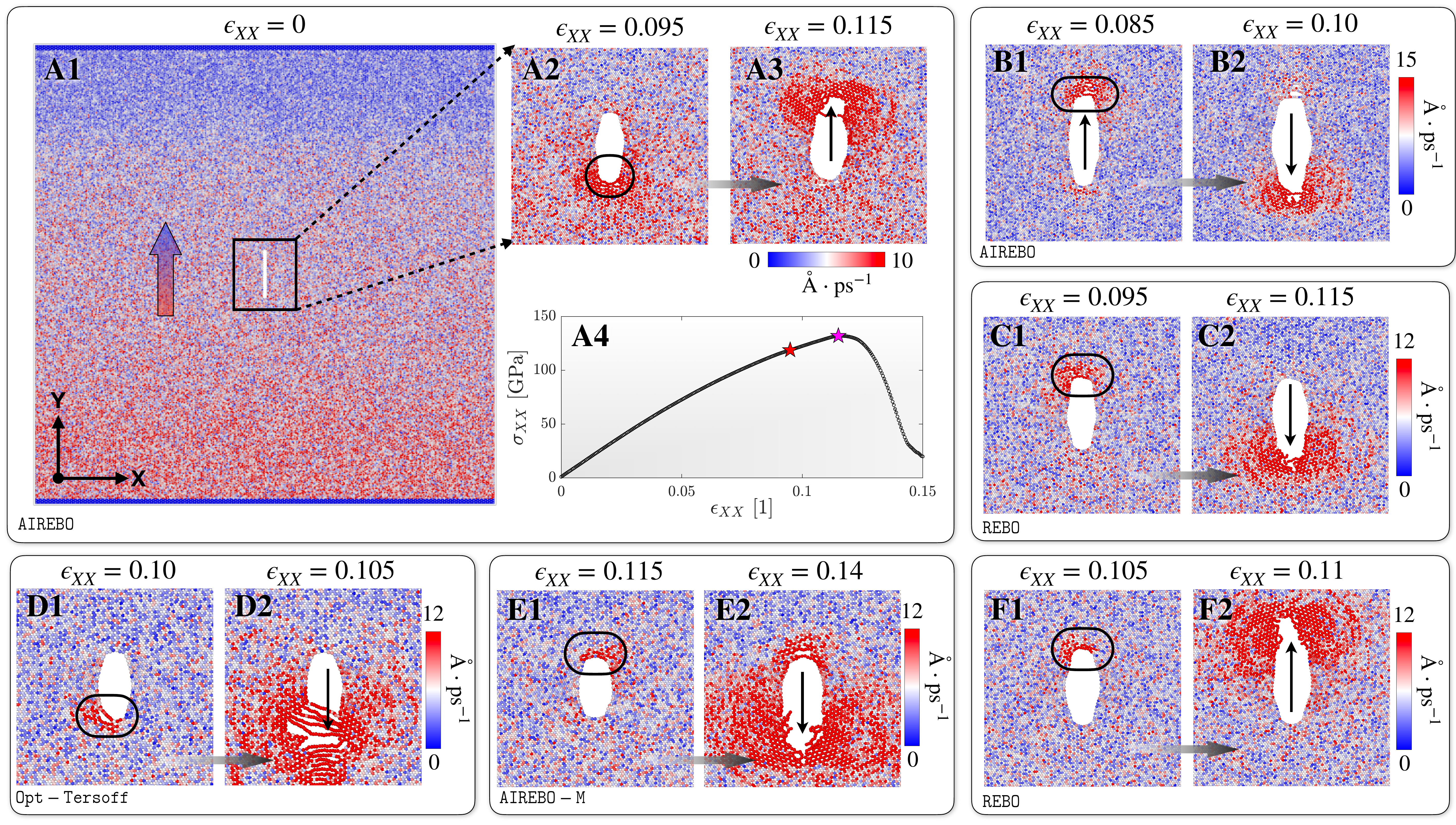}
    \caption{Illustration of anomalous fracture in ``REBO-based" potentials. The subfigures {\bf A} to {\bf F} are the 6 representative simulation cases of the fracture processes. {\bf A1} to {\bf A3} illustrate the anomalous fracture typically observed using the case of AIREBO potential under a 300 K temperature gradient as an example. The kinetic energy first accumulated on one end of the crack but then propagates on the other end instead. {\bf A4} is the corresponding stress-strain response. The red and pink stars correspond to {\bf A2} and {\bf A3}, respectively. {\bf B}. The anomalous fracture observed under a 0 K temperature gradient with the AIREBO potential. {\bf C}. The anomalous fracture observed with the REBO potential. {\bf D}. A normal fracture observed with the optimized Tersoff potential. {\bf E}. The anomalous fracture observed with the AIREBO-M potential. {\bf F}. A normal fracture observed with the REBO potential. The color bar indicates the atomic velocities in units of \AA$ \cdot$ ps$^{-1}$.}
    \label{frac_direc}
\end{figure}

By observing the crack profiles for all cases (Refer to Figures S6, S7, S8, and S9 in ESI for details), the initial directions of crack propagation are not guided by the temperature gradients but by an interesting phenomenon of kinetic energy transport along the crack tip is observed and characterized in Figure \ref{frac_direc}. The kinetic energy in the fracture is the difference between the released strain energy and the surface energy that drives the crack propagation, such that the propagation stops when the accumulated kinetic energy approaches zero \citep{kinetic_energy_fracture}. Intuitively, it may be expected that a crack will start propagating when sufficient kinetic energy accumulates on one side to drive the crack propagation. However, for the REBO, AIREBO, and AIREBO-M potentials, we unexpectedly observed that the kinetic energy accumulated on one end drove crack propagation on the other end instead. Figures \ref{frac_direc} \textbf{A}, \textbf{B}, \textbf{C}, \& \textbf{E} illustrate this phenomenon for AIREBO, REBO, AIREBO-M, under different temperature gradients. From Equations (\ref{rebo_eq}), (\ref{airebo_eq}), and (\ref{airebo-m_eq}), we attribute this anomalous fracture behavior observed in the ``REBO-based" potentials to the comparatively strong, short-range attractive forces that firmly resist crack propagation due to rapid accumulation of kinetic energy on one end, leading to the strain energy being released on other ends of the crack instead. Also, we believe the coupled thermal nonequilibrium and high strain rate loading also contribute to such ``blunt-liked'' mechanical responses. This anomalous fracture phenomenon is more frequently observed at higher temperature gradients and longer initial cracks, particularly for the AIREBO and AIREBO-M potentials (Refer to Table S2 in ESI). 

The visualizations indicate that the crack propagation does not strictly conform to brittle fracture: the crack does not propagate as a sharp notch like in Ref. \citep{nat_comm_KIC_smaller_value}, which adopted an initial defect length of 10 nm, which is longer than our longest initial defect length (8.1164 nm). Hu et al. demonstrated that porous graphene exhibited crack tip blunting behavior \citep{porous_graphene_blunting}. Hence, we believe that the short cracks in our simulations are analogous to a circular pore and the blunting behavior may be attributed to the higher width-to-length aspect ratio. The initial fracture of longer cracks is observed to be closer to pure brittle fractures (See Figures S6, S7, S8 of Section 6 in ESI). However, the optimized Tersoff model does not seem to be an ideal model for simulating the mechanical behaviors of graphene as they exhibit ``crazing-like" fractures (See Figure S9 in ESI), which should not be expected for graphene.

\subsection{Benchmarking Machine Learning Potentials\label{mlp_result_benchmark}}

\begin{figure}
    \centering
    \includegraphics[scale=0.25]{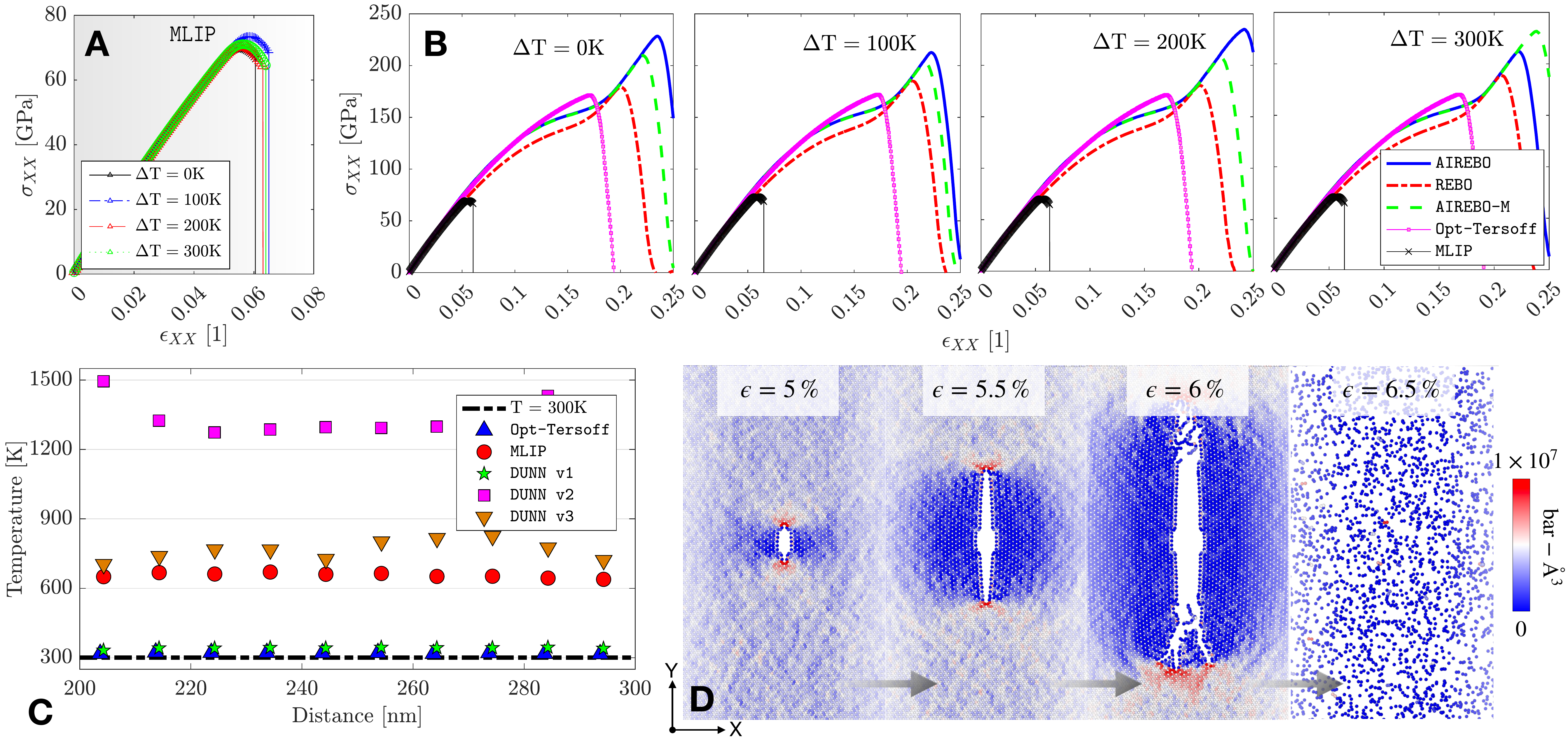}
    \caption{The simulation results for benchmarking the ML models. {\bf A}. The stress-strain responses of the graphene fracture with a small defect length ($L_C = 1.7217$ nm) using the MLIP potential, in which different temperature gradient environments are marked in different colors shown in the legend. {\bf B}. The stress-strain responses till fracture with the small initial defect, comparing the four empirical potentials with MLIP. Different colored lines represent the different potentials indicated in the legend. {\bf C}. The temperature distribution after 10,000 steps of equilibration run with preset 300 K constant temperature, where different colored and shaped dots represent the various MLPs benchmarked against the optimized Tersoff potential in blue triangular dots and the preset temperature is indicated as the dashed line at the bottom. {\bf D}. Graphical representation of a typical fracture process based on the MLIP potential at $\rm\Delta T = 300$ K, where the corresponding strains are marked on top. The virial atomic stresses are expressed in units of bar-\AA$^3$ (See color bar). }
    \label{ml_benchmark}
\end{figure}

We repeat the experiments in Section \ref{simulation_setup} to benchmark different MLPs (Figure \ref{ml_benchmark}). There are no clear relations that can be concluded between the mechanical responses to temperature gradients using the MLIP potential in Figure \ref{ml_benchmark} \textbf{A}, which agrees with the observation in Figure \ref{stress_strain}. For all four temperature gradients, Figure \ref{ml_benchmark} \textbf{B} suggests that the MLIP potential exhibits evidently smaller fracture stress compared to the four empirical potentials. If we compute the average value of the four empirical models under the four temperature gradients for benchmarking the MLIP fracture stress, we obtain ratios of $\sigma_\mathcal{F}^{\rm MLIP}\approx 0.39\sigma_\mathcal{F}^{\rm REBO}$, $\sigma_\mathcal{F}^{\rm MLIP}\approx 0.32\sigma_\mathcal{F}^{\rm AIREBO}$, $\sigma_\mathcal{F}^{\rm MLIP}\approx 0.33\sigma_\mathcal{F}^{\rm AIREBO-M}$. Figure \ref{ml_benchmark} \textbf{C} shows the thermal equilibration for the four MLP benchmarked against the optimized Tersoff potential which was specifically optimized to describe graphene's thermal properties more accurately than AIREBO and AIREBO-M \citep{potential_model}. Results in Figure \ref{ml_benchmark} \textbf{C} suggest that DUNN with a dropout rate of 0.1 (DUNN v1) has the most accurate temperature profile, as indicated by the green stars. Investigating the influence of the dropout ratio on the final configuration and various properties of graphene will be an interesting future direction. 

Figure \ref{ml_benchmark} \textbf{D} illustrates the fracture profile using the MLIP potential. The graphene layer fractures in very similar manners under different temperature gradients hence we illustrate the representative case of $\Delta \rm T=300\ K$. The fracture occurred at a strain of $\epsilon = 5\%$, which is $\frac{1}{3}$ the value of the empirical potentials with a similar initial defect length (Refer to ESI for the initial fracture profile). Brittle fracture is observed which matches experimental results \citep{brittle_exp_obser} and MD simulations \citep{nat_comm_KIC_smaller_value}. Immediately after fracture, the graphene sheet ``exploded" into scattered carbon atoms when $\epsilon=6.5\%$ (Figure \ref{ml_benchmark} \textbf{D}). This result indicates that the MLIP potential lacks the ability to model post-fracture deformations, which can possibly be attributed to the lack of relevant training data while constructing the model.

\section{Conclusion and outlook\label{conclusion}}

 Using the LAMMPS software \citep{lammps}, we performed non-equilibrium MD to study the fracture behavior of single-layer graphene sheets subjected to thermal gradients and compared the behavior in different potential models and varying sizes of initial defect lengths. We are essentially interested in answering three main questions: (1) What is the effect of the potential field on simulating the thermo-mechanical behavior of graphene? (2) What are the underlying mechanisms of fracture under extreme environments, i.e., high strain rates coupled with thermal gradients? What's more, the effects of different computational models, i.e., interatomic potentials are also of significance in this question. (3) What are the initial fracture characteristics and whether they are influenced by the thermal gradients and potential fields? We also determined whether non-equilibrium simulations at the molecular scale could match the theory of fracture mechanics. We further applied state-of-the-art \textit{ab initio} based MLPs to benchmark our simulations with empirical potentials and discussed the characteristics of the MLPs. These investigations provide a unique multiscale perspective from the first principle, to the molecular, to the sub-continuum scale.

We found that (1) for the AIREBO and AIREBO-M potentials, the fracture stresses were not positively correlated with the initial defect size; (2) strain-hardening effects were observed for ``REBO-based" potentials; (3) temperature gradients did not have direct effects on the mechanical responses; and (4) the AIREBO and AIREBO-M potentials exhibited higher fracture stresses compared to the optimized Tersoff and REBO potentials. For (2), we attributed the strain-hardening to coupled effects from the high strain rate, interatomic attractions, and transverse bond forces. We validated our MD simulations with QFM. We provided a novel perspective of nonlinear increases in fracture stresses at smaller initial crack lengths as transverse bonds helped to distribute the loads. Moreover, the AIREBO and AIREBO-M potentials were estimated to exhibit stronger attraction, leading to higher bond forces based on the observations of wider crack morphologies preceding fracture. The fracture stresses and strains were found to be unrelated to temperature gradients, and similarly so for the initial fracture directions and propagation. Interestingly, an anomalous fracture along the crack tips was observed and we explained that the short-ranged attractive forces in ``REBO-based'' potentials, coupled with high strain rates and thermal gradients attributes to this ``anomalous blunt-liked'' phenomena. Based on our results, the optimized Tersoff model could not capture graphene's fracture behaviors. We also proposed that the blunting effects of the short cracks might be attributed to the strain-hardening effects observed previously. The MLIP potential displayed smaller fracture stresses, approximately $\frac{1}{3}$ the values of the empirical potentials. With MLIP, brittle fractures were observed, but the atoms ``exploded" right after fracture which was unrealistic behavior. By benchmarking the four MLPs, the DUNN with a dropout rate of 0.1 had more ideal temperature distributions when compared with the optimized Tersoff potential.

In brief, we investigated the multiscale, thermo-mechanical coupling mechanism of graphene fracture under thermal gradients using computational MD simulations, theoretical fracture mechanics, and machine-learned models. Our study fills the gaps in (1) characterizing graphene's extreme mechanical behavior under heat gradients in non-equilibrium conditions and (2) benchmarking different machine-learned potentials for molecular simulations. This work can potentially inspire further studies and guide general industrial applications, such as fatigue and damage in graphene-based batteries and biosensors.

\section*{Data Availability}

The data and code used in this paper are available upon reasonable request to the authors.

\section*{Acknowledgement}

J.Y. acknowledges support from the US National Science Foundation (Grant Nos. 2038057 and 2223785). The authors also acknowledge computational resources provided by the XSEDE program under Grant TG-BIO210063.
\bibliography{sorsamp}

\providecommand{\noopsort}[1]{}\providecommand{\singleletter}[1]{#1}%
\begin{thebibliography}{90}%
\makeatletter
\providecommand \@ifxundefined [1]{%
 \@ifx{#1\undefined}
}%
\providecommand \@ifnum [1]{%
 \ifnum #1\expandafter \@firstoftwo
 \else \expandafter \@secondoftwo
 \fi
}%
\providecommand \@ifx [1]{%
 \ifx #1\expandafter \@firstoftwo
 \else \expandafter \@secondoftwo
 \fi
}%
\providecommand \natexlab [1]{#1}%
\providecommand \enquote  [1]{``#1''}%
\providecommand \bibnamefont  [1]{#1}%
\providecommand \bibfnamefont [1]{#1}%
\providecommand \citenamefont [1]{#1}%
\providecommand \href@noop [0]{\@secondoftwo}%
\providecommand \href [0]{\begingroup \@sanitize@url \@href}%
\providecommand \@href[1]{\@@startlink{#1}\@@href}%
\providecommand \@@href[1]{\endgroup#1\@@endlink}%
\providecommand \@sanitize@url [0]{\catcode `\\12\catcode `\$12\catcode
  `\&12\catcode `\#12\catcode `\^12\catcode `\_12\catcode `\%12\relax}%
\providecommand \@@startlink[1]{}%
\providecommand \@@endlink[0]{}%
\providecommand \url  [0]{\begingroup\@sanitize@url \@url }%
\providecommand \@url [1]{\endgroup\@href {#1}{\urlprefix }}%
\providecommand \urlprefix  [0]{URL }%
\providecommand \Eprint [0]{\href }%
\providecommand \doibase [0]{https://doi.org/}%
\providecommand \selectlanguage [0]{\@gobble}%
\providecommand \bibinfo  [0]{\@secondoftwo}%
\providecommand \bibfield  [0]{\@secondoftwo}%
\providecommand \translation [1]{[#1]}%
\providecommand \BibitemOpen [0]{}%
\providecommand \bibitemStop [0]{}%
\providecommand \bibitemNoStop [0]{.\EOS\space}%
\providecommand \EOS [0]{\spacefactor3000\relax}%
\providecommand \BibitemShut  [1]{\csname bibitem#1\endcsname}%
\let\auto@bib@innerbib\@empty
\bibitem [{\citenamefont {Akinwande}\ \emph {et~al.}(2017)\citenamefont
  {Akinwande}, \citenamefont {Brennan}, \citenamefont {Bunch}, \citenamefont
  {Egberts}, \citenamefont {Felts}, \citenamefont {Gao}, \citenamefont {Huang},
  \citenamefont {Kim}, \citenamefont {Li}, \citenamefont {Li}, \citenamefont
  {Liechti}, \citenamefont {Lu}, \citenamefont {Park}, \citenamefont {Reed},
  \citenamefont {Wang}, \citenamefont {Yakobson}, \citenamefont {Zhang},
  \citenamefont {Zhang}, \citenamefont {Zhou},\ and\ \citenamefont
  {Zhu}}]{graphene_mechanical}%
  \BibitemOpen
  \bibfield  {author} {\bibinfo {author} {\bibnamefont {Akinwande},
  \bibfnamefont {D.}}, \bibinfo {author} {\bibnamefont {Brennan}, \bibfnamefont
  {C.~J.}}, \bibinfo {author} {\bibnamefont {Bunch}, \bibfnamefont {J.~S.}},
  \bibinfo {author} {\bibnamefont {Egberts}, \bibfnamefont {P.}}, \bibinfo
  {author} {\bibnamefont {Felts}, \bibfnamefont {J.~R.}}, \bibinfo {author}
  {\bibnamefont {Gao}, \bibfnamefont {H.}}, \bibinfo {author} {\bibnamefont
  {Huang}, \bibfnamefont {R.}}, \bibinfo {author} {\bibnamefont {Kim},
  \bibfnamefont {J.-S.}}, \bibinfo {author} {\bibnamefont {Li}, \bibfnamefont
  {T.}}, \bibinfo {author} {\bibnamefont {Li}, \bibfnamefont {Y.}}, \bibinfo
  {author} {\bibnamefont {Liechti}, \bibfnamefont {K.~M.}}, \bibinfo {author}
  {\bibnamefont {Lu}, \bibfnamefont {N.}}, \bibinfo {author} {\bibnamefont
  {Park}, \bibfnamefont {H.~S.}}, \bibinfo {author} {\bibnamefont {Reed},
  \bibfnamefont {E.~J.}}, \bibinfo {author} {\bibnamefont {Wang}, \bibfnamefont
  {P.}}, \bibinfo {author} {\bibnamefont {Yakobson}, \bibfnamefont {B.~I.}},
  \bibinfo {author} {\bibnamefont {Zhang}, \bibfnamefont {T.}}, \bibinfo
  {author} {\bibnamefont {Zhang}, \bibfnamefont {Y.-W.}}, \bibinfo {author}
  {\bibnamefont {Zhou}, \bibfnamefont {Y.}}, and\ \bibinfo {author}
  {\bibnamefont {Zhu}, \bibfnamefont {Y.}},\ }\bibfield  {title} {\enquote
  {\bibinfo {title} {A review on mechanics and mechanical properties of 2d
  materials{\textemdash}graphene and beyond},}\ }\href@noop {} {\bibfield
  {journal} {\bibinfo  {journal} {Extreme Mechanics Letters}\ }\textbf
  {\bibinfo {volume} {13}},\ \bibinfo {pages} {42--77} (\bibinfo {year}
  {2017})}\BibitemShut {NoStop}%
\bibitem [{\citenamefont {Ando}(2009)}]{electric_1}%
  \BibitemOpen
  \bibfield  {author} {\bibinfo {author} {\bibnamefont {Ando}, \bibfnamefont
  {T.}},\ }\bibfield  {title} {\enquote {\bibinfo {title} {The electronic
  properties of graphene and carbon nanotubes},}\ }\href@noop {} {\bibfield
  {journal} {\bibinfo  {journal} {{NPG} Asia Materials}\ }\textbf {\bibinfo
  {volume} {1}},\ \bibinfo {pages} {17--21} (\bibinfo {year}
  {2009})}\BibitemShut {NoStop}%
\bibitem [{\citenamefont {Araujo}, \citenamefont {Terrones},\ and\
  \citenamefont {Dresselhaus}(2012)}]{Araujo2012}%
  \BibitemOpen
  \bibfield  {author} {\bibinfo {author} {\bibnamefont {Araujo}, \bibfnamefont
  {P.~T.}}, \bibinfo {author} {\bibnamefont {Terrones}, \bibfnamefont {M.}},
  and\ \bibinfo {author} {\bibnamefont {Dresselhaus}, \bibfnamefont {M.~S.}},\
  }\bibfield  {title} {\enquote {\bibinfo {title} {Defects and impurities in
  graphene-like materials},}\ }\href@noop {} {\bibfield  {journal} {\bibinfo
  {journal} {Materials Today}\ }\textbf {\bibinfo {volume} {15}},\ \bibinfo
  {pages} {98--109} (\bibinfo {year} {2012})}\BibitemShut {NoStop}%
\bibitem [{\citenamefont {Artrith}\ and\ \citenamefont
  {Behler}(2012)}]{behler2}%
  \BibitemOpen
  \bibfield  {author} {\bibinfo {author} {\bibnamefont {Artrith}, \bibfnamefont
  {N.}}and\ \bibinfo {author} {\bibnamefont {Behler}, \bibfnamefont {J.}},\
  }\bibfield  {title} {\enquote {\bibinfo {title} {High-dimensional neural
  network potentials for metal surfaces: A prototype study for copper},}\
  }\href {https://doi.org/10.1103/physrevb.85.045439} {\bibfield  {journal}
  {\bibinfo  {journal} {Physical Review B}\ }\textbf {\bibinfo {volume} {85}}
  (\bibinfo {year} {2012}),\ 10.1103/physrevb.85.045439}\BibitemShut {NoStop}%
\bibitem [{\citenamefont {Artrith}, \citenamefont {Morawietz},\ and\
  \citenamefont {Behler}(2011)}]{behler1}%
  \BibitemOpen
  \bibfield  {author} {\bibinfo {author} {\bibnamefont {Artrith}, \bibfnamefont
  {N.}}, \bibinfo {author} {\bibnamefont {Morawietz}, \bibfnamefont {T.}}, and\
  \bibinfo {author} {\bibnamefont {Behler}, \bibfnamefont {J.}},\ }\bibfield
  {title} {\enquote {\bibinfo {title} {High-dimensional neural-network
  potentials for multicomponent systems: Applications to zinc oxide},}\ }\href
  {https://doi.org/10.1103/physrevb.83.153101} {\bibfield  {journal} {\bibinfo
  {journal} {Physical Review B}\ }\textbf {\bibinfo {volume} {83}} (\bibinfo
  {year} {2011}),\ 10.1103/physrevb.83.153101}\BibitemShut {NoStop}%
\bibitem [{\citenamefont {Bagri}\ \emph {et~al.}(2011)\citenamefont {Bagri},
  \citenamefont {Kim}, \citenamefont {Ruoff},\ and\ \citenamefont
  {Shenoy}}]{defect_therm}%
  \BibitemOpen
  \bibfield  {author} {\bibinfo {author} {\bibnamefont {Bagri}, \bibfnamefont
  {A.}}, \bibinfo {author} {\bibnamefont {Kim}, \bibfnamefont {S.-P.}},
  \bibinfo {author} {\bibnamefont {Ruoff}, \bibfnamefont {R.~S.}}, and\
  \bibinfo {author} {\bibnamefont {Shenoy}, \bibfnamefont {V.~B.}},\ }\bibfield
   {title} {\enquote {\bibinfo {title} {Thermal transport across twin grain
  boundaries in polycrystalline graphene from nonequilibrium molecular dynamics
  simulations},}\ }\href@noop {} {\bibfield  {journal} {\bibinfo  {journal}
  {Nano Letters}\ }\textbf {\bibinfo {volume} {11}},\ \bibinfo {pages}
  {3917--3921} (\bibinfo {year} {2011})}\BibitemShut {NoStop}%
\bibitem [{\citenamefont {Balandin}(2011)}]{thermal_3}%
  \BibitemOpen
  \bibfield  {author} {\bibinfo {author} {\bibnamefont {Balandin},
  \bibfnamefont {A.~A.}},\ }\bibfield  {title} {\enquote {\bibinfo {title}
  {Thermal properties of graphene and nanostructured carbon materials},}\
  }\href@noop {} {\bibfield  {journal} {\bibinfo  {journal} {Nature Materials}\
  }\textbf {\bibinfo {volume} {10}},\ \bibinfo {pages} {569--581} (\bibinfo
  {year} {2011})}\BibitemShut {NoStop}%
\bibitem [{\citenamefont {Balandin}\ \emph {et~al.}(2008)\citenamefont
  {Balandin}, \citenamefont {Ghosh}, \citenamefont {Bao}, \citenamefont
  {Calizo}, \citenamefont {Teweldebrhan}, \citenamefont {Miao},\ and\
  \citenamefont {Lau}}]{thermal1}%
  \BibitemOpen
  \bibfield  {author} {\bibinfo {author} {\bibnamefont {Balandin},
  \bibfnamefont {A.~A.}}, \bibinfo {author} {\bibnamefont {Ghosh},
  \bibfnamefont {S.}}, \bibinfo {author} {\bibnamefont {Bao}, \bibfnamefont
  {W.}}, \bibinfo {author} {\bibnamefont {Calizo}, \bibfnamefont {I.}},
  \bibinfo {author} {\bibnamefont {Teweldebrhan}, \bibfnamefont {D.}}, \bibinfo
  {author} {\bibnamefont {Miao}, \bibfnamefont {F.}}, and\ \bibinfo {author}
  {\bibnamefont {Lau}, \bibfnamefont {C.~N.}},\ }\bibfield  {title} {\enquote
  {\bibinfo {title} {Superior thermal conductivity of single-layer graphene},}\
  }\href@noop {} {\bibfield  {journal} {\bibinfo  {journal} {Nano Letters}\
  }\textbf {\bibinfo {volume} {8}},\ \bibinfo {pages} {902--907} (\bibinfo
  {year} {2008})}\BibitemShut {NoStop}%
\bibitem [{\citenamefont {Behler}\ and\ \citenamefont
  {Parrinello}(2007)}]{Behler_PRL}%
  \BibitemOpen
  \bibfield  {author} {\bibinfo {author} {\bibnamefont {Behler}, \bibfnamefont
  {J.}}and\ \bibinfo {author} {\bibnamefont {Parrinello}, \bibfnamefont {M.}},\
  }\bibfield  {title} {\enquote {\bibinfo {title} {Generalized neural-network
  representation of high-dimensional potential-energy surfaces},}\ }\href
  {https://doi.org/10.1103/physrevlett.98.146401} {\bibfield  {journal}
  {\bibinfo  {journal} {Physical Review Letters}\ }\textbf {\bibinfo {volume}
  {98}} (\bibinfo {year} {2007}),\ 10.1103/physrevlett.98.146401}\BibitemShut
  {NoStop}%
\bibitem [{\citenamefont {Boretti}\ \emph {et~al.}(2018)\citenamefont
  {Boretti}, \citenamefont {Al-Zubaidy}, \citenamefont {Vaclavikova},
  \citenamefont {Al-Abri}, \citenamefont {Castelletto},\ and\ \citenamefont
  {Mikhalovsky}}]{water_1}%
  \BibitemOpen
  \bibfield  {author} {\bibinfo {author} {\bibnamefont {Boretti}, \bibfnamefont
  {A.}}, \bibinfo {author} {\bibnamefont {Al-Zubaidy}, \bibfnamefont {S.}},
  \bibinfo {author} {\bibnamefont {Vaclavikova}, \bibfnamefont {M.}}, \bibinfo
  {author} {\bibnamefont {Al-Abri}, \bibfnamefont {M.}}, \bibinfo {author}
  {\bibnamefont {Castelletto}, \bibfnamefont {S.}}, and\ \bibinfo {author}
  {\bibnamefont {Mikhalovsky}, \bibfnamefont {S.}},\ }\bibfield  {title}
  {\enquote {\bibinfo {title} {Outlook for graphene-based desalination
  membranes},}\ }\href@noop {} {\bibfield  {journal} {\bibinfo  {journal} {npj
  Clean Water}\ }\textbf {\bibinfo {volume} {1}} (\bibinfo {year}
  {2018})}\BibitemShut {NoStop}%
\bibitem [{\citenamefont {Brenner}(1990)}]{Brenner1990}%
  \BibitemOpen
  \bibfield  {author} {\bibinfo {author} {\bibnamefont {Brenner}, \bibfnamefont
  {D.~W.}},\ }\bibfield  {title} {\enquote {\bibinfo {title} {Empirical
  potential for hydrocarbons for use in simulating the chemical vapor
  deposition of diamond films},}\ }\href
  {https://doi.org/10.1103/physrevb.42.9458} {\bibfield  {journal} {\bibinfo
  {journal} {Physical Review B}\ }\textbf {\bibinfo {volume} {42}},\ \bibinfo
  {pages} {9458--9471} (\bibinfo {year} {1990})}\BibitemShut {NoStop}%
\bibitem [{\citenamefont {Brenner}(1992)}]{Brenner1992}%
  \BibitemOpen
  \bibfield  {author} {\bibinfo {author} {\bibnamefont {Brenner}, \bibfnamefont
  {D.~W.}},\ }\bibfield  {title} {\enquote {\bibinfo {title} {Erratum:
  Empirical potential for hydrocarbons for use in simulating the chemical vapor
  deposition of diamond films},}\ }\href
  {https://doi.org/10.1103/physrevb.46.1948.2} {\bibfield  {journal} {\bibinfo
  {journal} {Physical Review B}\ }\textbf {\bibinfo {volume} {46}},\ \bibinfo
  {pages} {1948--1948} (\bibinfo {year} {1992})}\BibitemShut {NoStop}%
\bibitem [{\citenamefont {Brenner}\ \emph {et~al.}(2002)\citenamefont
  {Brenner}, \citenamefont {Shenderova}, \citenamefont {Harrison},
  \citenamefont {Stuart}, \citenamefont {Ni},\ and\ \citenamefont
  {Sinnott}}]{rebo}%
  \BibitemOpen
  \bibfield  {author} {\bibinfo {author} {\bibnamefont {Brenner}, \bibfnamefont
  {D.~W.}}, \bibinfo {author} {\bibnamefont {Shenderova}, \bibfnamefont
  {O.~A.}}, \bibinfo {author} {\bibnamefont {Harrison}, \bibfnamefont {J.~A.}},
  \bibinfo {author} {\bibnamefont {Stuart}, \bibfnamefont {S.~J.}}, \bibinfo
  {author} {\bibnamefont {Ni}, \bibfnamefont {B.}}, and\ \bibinfo {author}
  {\bibnamefont {Sinnott}, \bibfnamefont {S.~B.}},\ }\bibfield  {title}
  {\enquote {\bibinfo {title} {A second-generation reactive empirical bond
  order ({REBO}) potential energy expression for hydrocarbons},}\ }\href
  {https://doi.org/10.1088/0953-8984/14/4/312} {\bibfield  {journal} {\bibinfo
  {journal} {Journal of Physics: Condensed Matter}\ }\textbf {\bibinfo {volume}
  {14}},\ \bibinfo {pages} {783--802} (\bibinfo {year} {2002})}\BibitemShut
  {NoStop}%
\bibitem [{\citenamefont {Bu}\ \emph {et~al.}(2009)\citenamefont {Bu},
  \citenamefont {Chen}, \citenamefont {Zou}, \citenamefont {Yi}, \citenamefont
  {Bi},\ and\ \citenamefont {Ni}}]{phys_lett_a}%
  \BibitemOpen
  \bibfield  {author} {\bibinfo {author} {\bibnamefont {Bu}, \bibfnamefont
  {H.}}, \bibinfo {author} {\bibnamefont {Chen}, \bibfnamefont {Y.}}, \bibinfo
  {author} {\bibnamefont {Zou}, \bibfnamefont {M.}}, \bibinfo {author}
  {\bibnamefont {Yi}, \bibfnamefont {H.}}, \bibinfo {author} {\bibnamefont
  {Bi}, \bibfnamefont {K.}}, and\ \bibinfo {author} {\bibnamefont {Ni},
  \bibfnamefont {Z.}},\ }\bibfield  {title} {\enquote {\bibinfo {title}
  {Atomistic simulations of mechanical properties of graphene nanoribbons},}\
  }\href {https://doi.org/10.1016/j.physleta.2009.07.048} {\bibfield  {journal}
  {\bibinfo  {journal} {Physics Letters A}\ }\textbf {\bibinfo {volume}
  {373}},\ \bibinfo {pages} {3359--3362} (\bibinfo {year} {2009})}\BibitemShut
  {NoStop}%
\bibitem [{\citenamefont {Bunch}(2008)}]{cornell_thesis_2008}%
  \BibitemOpen
  \bibfield  {author} {\bibinfo {author} {\bibnamefont {Bunch}, \bibfnamefont
  {J.~S.}},\ }\href@noop {} {\emph {\bibinfo {title} {Mechanical and electrical
  properties of graphene sheets}}}\ (\bibinfo  {publisher} {Citeseer},\
  \bibinfo {year} {2008})\BibitemShut {NoStop}%
\bibitem [{\citenamefont {Chen}\ \emph {et~al.}(2017)\citenamefont {Chen},
  \citenamefont {Xu}, \citenamefont {Wang}, \citenamefont {Huang},
  \citenamefont {Xi}, \citenamefont {Cai}, \citenamefont {Guo}, \citenamefont
  {Xu}, \citenamefont {Gao},\ and\ \citenamefont {Gao}}]{graphene_battery_2}%
  \BibitemOpen
  \bibfield  {author} {\bibinfo {author} {\bibnamefont {Chen}, \bibfnamefont
  {H.}}, \bibinfo {author} {\bibnamefont {Xu}, \bibfnamefont {H.}}, \bibinfo
  {author} {\bibnamefont {Wang}, \bibfnamefont {S.}}, \bibinfo {author}
  {\bibnamefont {Huang}, \bibfnamefont {T.}}, \bibinfo {author} {\bibnamefont
  {Xi}, \bibfnamefont {J.}}, \bibinfo {author} {\bibnamefont {Cai},
  \bibfnamefont {S.}}, \bibinfo {author} {\bibnamefont {Guo}, \bibfnamefont
  {F.}}, \bibinfo {author} {\bibnamefont {Xu}, \bibfnamefont {Z.}}, \bibinfo
  {author} {\bibnamefont {Gao}, \bibfnamefont {W.}}, and\ \bibinfo {author}
  {\bibnamefont {Gao}, \bibfnamefont {C.}},\ }\bibfield  {title} {\enquote
  {\bibinfo {title} {Ultrafast all-climate aluminum-graphene battery with
  quarter-million cycle life},}\ }\href@noop {} {\bibfield  {journal} {\bibinfo
   {journal} {Science Advances}\ }\textbf {\bibinfo {volume} {3}} (\bibinfo
  {year} {2017})}\BibitemShut {NoStop}%
\bibitem [{\citenamefont {Cohen-Tanugi}\ and\ \citenamefont
  {Grossman}(2014)}]{grossman}%
  \BibitemOpen
  \bibfield  {author} {\bibinfo {author} {\bibnamefont {Cohen-Tanugi},
  \bibfnamefont {D.}}and\ \bibinfo {author} {\bibnamefont {Grossman},
  \bibfnamefont {J.~C.}},\ }\bibfield  {title} {\enquote {\bibinfo {title}
  {Mechanical strength of nanoporous graphene as a desalination membrane},}\
  }\href {https://doi.org/10.1021/nl502399y} {\bibfield  {journal} {\bibinfo
  {journal} {Nano Letters}\ }\textbf {\bibinfo {volume} {14}},\ \bibinfo
  {pages} {6171--6178} (\bibinfo {year} {2014})}\BibitemShut {NoStop}%
\bibitem [{\citenamefont {Creager}\ and\ \citenamefont
  {Paris}(1967)}]{frac_2_deriv}%
  \BibitemOpen
  \bibfield  {author} {\bibinfo {author} {\bibnamefont {Creager}, \bibfnamefont
  {M.}}and\ \bibinfo {author} {\bibnamefont {Paris}, \bibfnamefont {P.~C.}},\
  }\bibfield  {title} {\enquote {\bibinfo {title} {Elastic field equations for
  blunt cracks with reference to stress corrosion cracking},}\ }\href
  {https://doi.org/10.1007/bf00182890} {\bibfield  {journal} {\bibinfo
  {journal} {International Journal of Fracture Mechanics}\ }\textbf {\bibinfo
  {volume} {3}},\ \bibinfo {pages} {247--252} (\bibinfo {year}
  {1967})}\BibitemShut {NoStop}%
\bibitem [{\citenamefont {David}(2003)}]{morse}%
  \BibitemOpen
  \bibfield  {author} {\bibinfo {author} {\bibnamefont {David}, \bibfnamefont
  {C.}},\ }\href {https://chemphys.uconn.edu/~ch351vc/pdfs/morse.pdf} {\enquote
  {\bibinfo {title} {The morse potential},}\ } (\bibinfo {year} {2003}),\
  \bibinfo {note} {physical Chemistry 351 at the University of
  Connecticut}\BibitemShut {NoStop}%
\bibitem [{\citenamefont {Drory}\ \emph {et~al.}(1995)\citenamefont {Drory},
  \citenamefont {Dauskardt}, \citenamefont {Kant},\ and\ \citenamefont
  {Ritchie}}]{frac_1_deriv}%
  \BibitemOpen
  \bibfield  {author} {\bibinfo {author} {\bibnamefont {Drory}, \bibfnamefont
  {M.~D.}}, \bibinfo {author} {\bibnamefont {Dauskardt}, \bibfnamefont
  {R.~H.}}, \bibinfo {author} {\bibnamefont {Kant}, \bibfnamefont {A.}}, and\
  \bibinfo {author} {\bibnamefont {Ritchie}, \bibfnamefont {R.~O.}},\
  }\bibfield  {title} {\enquote {\bibinfo {title} {Fracture of synthetic
  diamond},}\ }\href {https://doi.org/10.1063/1.360060} {\bibfield  {journal}
  {\bibinfo  {journal} {Journal of Applied Physics}\ }\textbf {\bibinfo
  {volume} {78}},\ \bibinfo {pages} {3083--3088} (\bibinfo {year}
  {1995})}\BibitemShut {NoStop}%
\bibitem [{\citenamefont {Felix}\ \emph {et~al.}(2020)\citenamefont {Felix},
  \citenamefont {Tromer}, \citenamefont {Autreto}, \citenamefont {Junior},\
  and\ \citenamefont {Galvao}}]{mechanical_thermal_monolayer}%
  \BibitemOpen
  \bibfield  {author} {\bibinfo {author} {\bibnamefont {Felix}, \bibfnamefont
  {L.~C.}}, \bibinfo {author} {\bibnamefont {Tromer}, \bibfnamefont {R.~M.}},
  \bibinfo {author} {\bibnamefont {Autreto}, \bibfnamefont {P.~A.~S.}},
  \bibinfo {author} {\bibnamefont {Junior}, \bibfnamefont {L.~A.~R.}}, and\
  \bibinfo {author} {\bibnamefont {Galvao}, \bibfnamefont {D.~S.}},\ }\bibfield
   {title} {\enquote {\bibinfo {title} {On the mechanical properties and
  thermal stability of a recently synthesized monolayer amorphous carbon},}\
  }\href@noop {} {\bibfield  {journal} {\bibinfo  {journal} {The Journal of
  Physical Chemistry C}\ }\textbf {\bibinfo {volume} {124}},\ \bibinfo {pages}
  {14855--14860} (\bibinfo {year} {2020})}\BibitemShut {NoStop}%
\bibitem [{\citenamefont {Ferrante}, \citenamefont {Smith},\ and\ \citenamefont
  {Rose}(1983)}]{ferrante1983}%
  \BibitemOpen
  \bibfield  {author} {\bibinfo {author} {\bibnamefont {Ferrante},
  \bibfnamefont {J.}}, \bibinfo {author} {\bibnamefont {Smith}, \bibfnamefont
  {J.~R.}}, and\ \bibinfo {author} {\bibnamefont {Rose}, \bibfnamefont
  {J.~H.}},\ }\bibfield  {title} {\enquote {\bibinfo {title} {Diatomic
  molecules and metallic adhesion, cohesion, and chemisorption: A single
  binding-energy relation},}\ }\href@noop {} {\bibfield  {journal} {\bibinfo
  {journal} {Phys. Rev. Lett.}\ }\textbf {\bibinfo {volume} {50}},\ \bibinfo
  {pages} {1385--1386} (\bibinfo {year} {1983})}\BibitemShut {NoStop}%
\bibitem [{\citenamefont {Gao}\ \emph {et~al.}(2020)\citenamefont {Gao},
  \citenamefont {Ramezanghorbani}, \citenamefont {Isayev}, \citenamefont
  {Smith},\ and\ \citenamefont {Roitberg}}]{torchani}%
  \BibitemOpen
  \bibfield  {author} {\bibinfo {author} {\bibnamefont {Gao}, \bibfnamefont
  {X.}}, \bibinfo {author} {\bibnamefont {Ramezanghorbani}, \bibfnamefont
  {F.}}, \bibinfo {author} {\bibnamefont {Isayev}, \bibfnamefont {O.}},
  \bibinfo {author} {\bibnamefont {Smith}, \bibfnamefont {J.~S.}}, and\
  \bibinfo {author} {\bibnamefont {Roitberg}, \bibfnamefont {A.~E.}},\
  }\bibfield  {title} {\enquote {\bibinfo {title} {{TorchANI}: A free and open
  source {PyTorch}-based deep learning implementation of the {ANI} neural
  network potentials},}\ }\href {https://doi.org/10.1021/acs.jcim.0c00451}
  {\bibfield  {journal} {\bibinfo  {journal} {Journal of Chemical Information
  and Modeling}\ }\textbf {\bibinfo {volume} {60}},\ \bibinfo {pages}
  {3408--3415} (\bibinfo {year} {2020})}\BibitemShut {NoStop}%
\bibitem [{\citenamefont {Grantab}, \citenamefont {Shenoy},\ and\ \citenamefont
  {Ruoff}(2010)}]{defect_mech1}%
  \BibitemOpen
  \bibfield  {author} {\bibinfo {author} {\bibnamefont {Grantab}, \bibfnamefont
  {R.}}, \bibinfo {author} {\bibnamefont {Shenoy}, \bibfnamefont {V.~B.}}, and\
  \bibinfo {author} {\bibnamefont {Ruoff}, \bibfnamefont {R.~S.}},\ }\bibfield
  {title} {\enquote {\bibinfo {title} {Anomalous strength characteristics of
  tilt grain boundaries in graphene},}\ }\href@noop {} {\bibfield  {journal}
  {\bibinfo  {journal} {Science}\ }\textbf {\bibinfo {volume} {330}},\ \bibinfo
  {pages} {946--948} (\bibinfo {year} {2010})}\BibitemShut {NoStop}%
\bibitem [{\citenamefont {Gu}\ \emph {et~al.}(2018)\citenamefont {Gu},
  \citenamefont {Wei}, \citenamefont {Yin}, \citenamefont {Li},\ and\
  \citenamefont {Yang}}]{aps_thermal}%
  \BibitemOpen
  \bibfield  {author} {\bibinfo {author} {\bibnamefont {Gu}, \bibfnamefont
  {X.}}, \bibinfo {author} {\bibnamefont {Wei}, \bibfnamefont {Y.}}, \bibinfo
  {author} {\bibnamefont {Yin}, \bibfnamefont {X.}}, \bibinfo {author}
  {\bibnamefont {Li}, \bibfnamefont {B.}}, and\ \bibinfo {author} {\bibnamefont
  {Yang}, \bibfnamefont {R.}},\ }\bibfield  {title} {\enquote {\bibinfo {title}
  {Colloquium: Phononic thermal properties of two-dimensional materials},}\
  }\href@noop {} {\bibfield  {journal} {\bibinfo  {journal} {Rev. Mod. Phys.}\
  }\textbf {\bibinfo {volume} {90}},\ \bibinfo {pages} {041002} (\bibinfo
  {year} {2018})}\BibitemShut {NoStop}%
\bibitem [{\citenamefont {Guo}\ \emph {et~al.}(2015)\citenamefont {Guo},
  \citenamefont {Kitamura}, \citenamefont {Yan}, \citenamefont {Sumigawa},\
  and\ \citenamefont {Huang}}]{kinetic_energy_fracture}%
  \BibitemOpen
  \bibfield  {author} {\bibinfo {author} {\bibnamefont {Guo}, \bibfnamefont
  {L.}}, \bibinfo {author} {\bibnamefont {Kitamura}, \bibfnamefont {T.}},
  \bibinfo {author} {\bibnamefont {Yan}, \bibfnamefont {Y.}}, \bibinfo {author}
  {\bibnamefont {Sumigawa}, \bibfnamefont {T.}}, and\ \bibinfo {author}
  {\bibnamefont {Huang}, \bibfnamefont {K.}},\ }\bibfield  {title} {\enquote
  {\bibinfo {title} {Fracture mechanics investigation on crack propagation in
  the nano-multilayered materials},}\ }\href
  {https://doi.org/10.1016/j.ijsolstr.2015.03.025} {\bibfield  {journal}
  {\bibinfo  {journal} {International Journal of Solids and Structures}\
  }\textbf {\bibinfo {volume} {64-65}},\ \bibinfo {pages} {208--220} (\bibinfo
  {year} {2015})}\BibitemShut {NoStop}%
\bibitem [{\citenamefont {Hashimoto}\ \emph {et~al.}(2004)\citenamefont
  {Hashimoto}, \citenamefont {Suenaga}, \citenamefont {Gloter}, \citenamefont
  {Urita},\ and\ \citenamefont {Iijima}}]{defect_evidence}%
  \BibitemOpen
  \bibfield  {author} {\bibinfo {author} {\bibnamefont {Hashimoto},
  \bibfnamefont {A.}}, \bibinfo {author} {\bibnamefont {Suenaga}, \bibfnamefont
  {K.}}, \bibinfo {author} {\bibnamefont {Gloter}, \bibfnamefont {A.}},
  \bibinfo {author} {\bibnamefont {Urita}, \bibfnamefont {K.}}, and\ \bibinfo
  {author} {\bibnamefont {Iijima}, \bibfnamefont {S.}},\ }\bibfield  {title}
  {\enquote {\bibinfo {title} {Direct evidence for atomic defects in graphene
  layers},}\ }\href@noop {} {\bibfield  {journal} {\bibinfo  {journal}
  {Nature}\ }\textbf {\bibinfo {volume} {430}},\ \bibinfo {pages} {870--873}
  (\bibinfo {year} {2004})}\BibitemShut {NoStop}%
\bibitem [{\citenamefont {Homaeigohar}\ and\ \citenamefont
  {Elbahri}(2017)}]{water_2}%
  \BibitemOpen
  \bibfield  {author} {\bibinfo {author} {\bibnamefont {Homaeigohar},
  \bibfnamefont {S.}}and\ \bibinfo {author} {\bibnamefont {Elbahri},
  \bibfnamefont {M.}},\ }\bibfield  {title} {\enquote {\bibinfo {title}
  {Graphene membranes for water desalination},}\ }\href@noop {} {\bibfield
  {journal} {\bibinfo  {journal} {{NPG} Asia Materials}\ }\textbf {\bibinfo
  {volume} {9}},\ \bibinfo {pages} {e427--e427} (\bibinfo {year}
  {2017})}\BibitemShut {NoStop}%
\bibitem [{\citenamefont {Hu}, \citenamefont {Ruan},\ and\ \citenamefont
  {Chen}(2009)}]{thermal2}%
  \BibitemOpen
  \bibfield  {author} {\bibinfo {author} {\bibnamefont {Hu}, \bibfnamefont
  {J.}}, \bibinfo {author} {\bibnamefont {Ruan}, \bibfnamefont {X.}}, and\
  \bibinfo {author} {\bibnamefont {Chen}, \bibfnamefont {Y.~P.}},\ }\bibfield
  {title} {\enquote {\bibinfo {title} {Thermal conductivity and thermal
  rectification in graphene nanoribbons: A molecular dynamics study},}\
  }\href@noop {} {\bibfield  {journal} {\bibinfo  {journal} {Nano Letters}\
  }\textbf {\bibinfo {volume} {9}},\ \bibinfo {pages} {2730--2735} (\bibinfo
  {year} {2009})}\BibitemShut {NoStop}%
\bibitem [{\citenamefont {Hu}\ \emph {et~al.}(2021)\citenamefont {Hu},
  \citenamefont {Zhou}, \citenamefont {Zhang}, \citenamefont {Yi},\ and\
  \citenamefont {Wang}}]{mechanical_temperature}%
  \BibitemOpen
  \bibfield  {author} {\bibinfo {author} {\bibnamefont {Hu}, \bibfnamefont
  {J.}}, \bibinfo {author} {\bibnamefont {Zhou}, \bibfnamefont {J.}}, \bibinfo
  {author} {\bibnamefont {Zhang}, \bibfnamefont {A.}}, \bibinfo {author}
  {\bibnamefont {Yi}, \bibfnamefont {L.}}, and\ \bibinfo {author} {\bibnamefont
  {Wang}, \bibfnamefont {J.}},\ }\bibfield  {title} {\enquote {\bibinfo {title}
  {Temperature dependent mechanical properties of graphene based carbon
  honeycombs under tension and compression},}\ }\href@noop {} {\bibfield
  {journal} {\bibinfo  {journal} {Physics Letters A}\ }\textbf {\bibinfo
  {volume} {391}},\ \bibinfo {pages} {127130} (\bibinfo {year}
  {2021})}\BibitemShut {NoStop}%
\bibitem [{\citenamefont {Hu}\ \emph {et~al.}(2015)\citenamefont {Hu},
  \citenamefont {Wyant}, \citenamefont {Muniz}, \citenamefont
  {Ramasubramaniam},\ and\ \citenamefont
  {Maroudas}}]{porous_graphene_blunting}%
  \BibitemOpen
  \bibfield  {author} {\bibinfo {author} {\bibnamefont {Hu}, \bibfnamefont
  {L.}}, \bibinfo {author} {\bibnamefont {Wyant}, \bibfnamefont {S.}}, \bibinfo
  {author} {\bibnamefont {Muniz}, \bibfnamefont {A.~R.}}, \bibinfo {author}
  {\bibnamefont {Ramasubramaniam}, \bibfnamefont {A.}}, and\ \bibinfo {author}
  {\bibnamefont {Maroudas}, \bibfnamefont {D.}},\ }\bibfield  {title} {\enquote
  {\bibinfo {title} {Mechanical behavior and fracture of graphene
  nanomeshes},}\ }\href {https://doi.org/10.1063/1.4905583} {\bibfield
  {journal} {\bibinfo  {journal} {Journal of Applied Physics}\ }\textbf
  {\bibinfo {volume} {117}},\ \bibinfo {pages} {024302} (\bibinfo {year}
  {2015})}\BibitemShut {NoStop}%
\bibitem [{\citenamefont {Hwangbo}\ \emph {et~al.}(2014)\citenamefont
  {Hwangbo}, \citenamefont {Lee}, \citenamefont {Kim}, \citenamefont {Kim},
  \citenamefont {Kim}, \citenamefont {Jang}, \citenamefont {Lee}, \citenamefont
  {Lee}, \citenamefont {Kim}, \citenamefont {Ahn},\ and\ \citenamefont
  {Lee}}]{fracture_toughness_10}%
  \BibitemOpen
  \bibfield  {author} {\bibinfo {author} {\bibnamefont {Hwangbo}, \bibfnamefont
  {Y.}}, \bibinfo {author} {\bibnamefont {Lee}, \bibfnamefont {C.-K.}},
  \bibinfo {author} {\bibnamefont {Kim}, \bibfnamefont {S.-M.}}, \bibinfo
  {author} {\bibnamefont {Kim}, \bibfnamefont {J.-H.}}, \bibinfo {author}
  {\bibnamefont {Kim}, \bibfnamefont {K.-S.}}, \bibinfo {author} {\bibnamefont
  {Jang}, \bibfnamefont {B.}}, \bibinfo {author} {\bibnamefont {Lee},
  \bibfnamefont {H.-J.}}, \bibinfo {author} {\bibnamefont {Lee}, \bibfnamefont
  {S.-K.}}, \bibinfo {author} {\bibnamefont {Kim}, \bibfnamefont {S.-S.}},
  \bibinfo {author} {\bibnamefont {Ahn}, \bibfnamefont {J.-H.}}, and\ \bibinfo
  {author} {\bibnamefont {Lee}, \bibfnamefont {S.-M.}},\ }\bibfield  {title}
  {\enquote {\bibinfo {title} {Fracture characteristics of monolayer
  {CVD}-graphene},}\ }\href {https://doi.org/10.1038/srep04439} {\bibfield
  {journal} {\bibinfo  {journal} {Scientific Reports}\ }\textbf {\bibinfo
  {volume} {4}} (\bibinfo {year} {2014}),\ 10.1038/srep04439}\BibitemShut
  {NoStop}%
\bibitem [{\citenamefont {Irwin}(1957)}]{fracture_criterion}%
  \BibitemOpen
  \bibfield  {author} {\bibinfo {author} {\bibnamefont {Irwin}, \bibfnamefont
  {G.~R.}},\ }\bibfield  {title} {\enquote {\bibinfo {title} {Analysis of
  stresses and strains near the end of a crack traversing a plate},}\ }\href
  {https://doi.org/10.1115/1.4011547} {\bibfield  {journal} {\bibinfo
  {journal} {Journal of Applied Mechanics}\ }\textbf {\bibinfo {volume} {24}},\
  \bibinfo {pages} {361--364} (\bibinfo {year} {1957})}\BibitemShut {NoStop}%
\bibitem [{\citenamefont {Jangid}\ and\ \citenamefont
  {Kottantharayil}(2020)}]{heat_reconstruct}%
  \BibitemOpen
  \bibfield  {author} {\bibinfo {author} {\bibnamefont {Jangid}, \bibfnamefont
  {P.}}and\ \bibinfo {author} {\bibnamefont {Kottantharayil}, \bibfnamefont
  {A.}},\ }\bibfield  {title} {\enquote {\bibinfo {title} {Reconstruction of
  fractured graphene by thermal treatment in methane gas},}\ }\href@noop {}
  {\bibfield  {journal} {\bibinfo  {journal} {Materials Science and
  Engineering: B}\ }\textbf {\bibinfo {volume} {260}},\ \bibinfo {pages}
  {114625} (\bibinfo {year} {2020})}\BibitemShut {NoStop}%
\bibitem [{\citenamefont {Jung}, \citenamefont {Myung},\ and\ \citenamefont
  {Irle}(2022)}]{mlp_graphene_gangseob}%
  \BibitemOpen
  \bibfield  {author} {\bibinfo {author} {\bibnamefont {Jung}, \bibfnamefont
  {G.~S.}}, \bibinfo {author} {\bibnamefont {Myung}, \bibfnamefont {H.~J.}},
  and\ \bibinfo {author} {\bibnamefont {Irle}, \bibfnamefont {S.}},\ }\href
  {https://doi.org/10.21203/rs.3.rs-1178290/v1} {\enquote {\bibinfo {title}
  {Artificial neural network potentials for mechanics and fracture dynamics of
  materials},}\ } (\bibinfo {year} {2022})\BibitemShut {NoStop}%
\bibitem [{\citenamefont {Jung}\ \emph {et~al.}(2017)\citenamefont {Jung},
  \citenamefont {Yeo}, \citenamefont {Tian}, \citenamefont {Qin},\ and\
  \citenamefont {Buehler}}]{jj_nanoscale}%
  \BibitemOpen
  \bibfield  {author} {\bibinfo {author} {\bibnamefont {Jung}, \bibfnamefont
  {G.~S.}}, \bibinfo {author} {\bibnamefont {Yeo}, \bibfnamefont {J.}},
  \bibinfo {author} {\bibnamefont {Tian}, \bibfnamefont {Z.}}, \bibinfo
  {author} {\bibnamefont {Qin}, \bibfnamefont {Z.}}, and\ \bibinfo {author}
  {\bibnamefont {Buehler}, \bibfnamefont {M.~J.}},\ }\bibfield  {title}
  {\enquote {\bibinfo {title} {Unusually low and density-insensitive thermal
  conductivity of three-dimensional gyroid graphene},}\ }\href@noop {}
  {\bibfield  {journal} {\bibinfo  {journal} {Nanoscale}\ }\textbf {\bibinfo
  {volume} {9}},\ \bibinfo {pages} {13477--13484} (\bibinfo {year}
  {2017})}\BibitemShut {NoStop}%
\bibitem [{\citenamefont {Kasirga}(2020)}]{thermal_book}%
  \BibitemOpen
  \bibfield  {author} {\bibinfo {author} {\bibnamefont {Kasirga}, \bibfnamefont
  {T.~S.}},\ }\bibfield  {title} {\enquote {\bibinfo {title} {Thermal
  conductivity measurements in~2d materials},}\ }in\ \href@noop {} {\emph
  {\bibinfo {booktitle} {Thermal Conductivity Measurements in Atomically Thin
  Materials and Devices}}}\ (\bibinfo  {publisher} {Springer Singapore},\
  \bibinfo {year} {2020})\ pp.\ \bibinfo {pages} {11--27}\BibitemShut {NoStop}%
\bibitem [{\citenamefont {Lee}\ \emph {et~al.}(2008)\citenamefont {Lee},
  \citenamefont {Wei}, \citenamefont {Kysar},\ and\ \citenamefont
  {Hone}}]{elastic2}%
  \BibitemOpen
  \bibfield  {author} {\bibinfo {author} {\bibnamefont {Lee}, \bibfnamefont
  {C.}}, \bibinfo {author} {\bibnamefont {Wei}, \bibfnamefont {X.}}, \bibinfo
  {author} {\bibnamefont {Kysar}, \bibfnamefont {J.~W.}}, and\ \bibinfo
  {author} {\bibnamefont {Hone}, \bibfnamefont {J.}},\ }\bibfield  {title}
  {\enquote {\bibinfo {title} {Measurement of the elastic properties and
  intrinsic strength of monolayer graphene},}\ }\href@noop {} {\bibfield
  {journal} {\bibinfo  {journal} {Science}\ }\textbf {\bibinfo {volume}
  {321}},\ \bibinfo {pages} {385--388} (\bibinfo {year} {2008})}\BibitemShut
  {NoStop}%
\bibitem [{\citenamefont {Li}\ \emph {et~al.}(2019)\citenamefont {Li},
  \citenamefont {Deng}, \citenamefont {Zheng}, \citenamefont {Zhang},
  \citenamefont {Liao},\ and\ \citenamefont
  {Zhou}}]{defect_mechanical_thermal}%
  \BibitemOpen
  \bibfield  {author} {\bibinfo {author} {\bibnamefont {Li}, \bibfnamefont
  {M.}}, \bibinfo {author} {\bibnamefont {Deng}, \bibfnamefont {T.}}, \bibinfo
  {author} {\bibnamefont {Zheng}, \bibfnamefont {B.}}, \bibinfo {author}
  {\bibnamefont {Zhang}, \bibfnamefont {Y.}}, \bibinfo {author} {\bibnamefont
  {Liao}, \bibfnamefont {Y.}}, and\ \bibinfo {author} {\bibnamefont {Zhou},
  \bibfnamefont {H.}},\ }\bibfield  {title} {\enquote {\bibinfo {title} {Effect
  of defects on the mechanical and thermal properties of graphene},}\
  }\href@noop {} {\bibfield  {journal} {\bibinfo  {journal} {Nanomaterials}\
  }\textbf {\bibinfo {volume} {9}} (\bibinfo {year} {2019})}\BibitemShut
  {NoStop}%
\bibitem [{\citenamefont {Lindsay}\ and\ \citenamefont
  {Broido}(2010)}]{opt-tersoff}%
  \BibitemOpen
  \bibfield  {author} {\bibinfo {author} {\bibnamefont {Lindsay}, \bibfnamefont
  {L.}}and\ \bibinfo {author} {\bibnamefont {Broido}, \bibfnamefont {D.~A.}},\
  }\bibfield  {title} {\enquote {\bibinfo {title} {Optimized tersoff and
  brenner empirical potential parameters for lattice dynamics and phonon
  thermal transport in carbon nanotubes and graphene},}\ }\href@noop {}
  {\bibfield  {journal} {\bibinfo  {journal} {Physical Review B}\ }\textbf
  {\bibinfo {volume} {81}} (\bibinfo {year} {2010})}\BibitemShut {NoStop}%
\bibitem [{\citenamefont {Liu}\ and\ \citenamefont
  {Wu}(2016)}]{jmr_graphene_mechanical}%
  \BibitemOpen
  \bibfield  {author} {\bibinfo {author} {\bibnamefont {Liu}, \bibfnamefont
  {K.}}and\ \bibinfo {author} {\bibnamefont {Wu}, \bibfnamefont {J.}},\
  }\bibfield  {title} {\enquote {\bibinfo {title} {Mechanical properties of
  two-dimensional materials and heterostructures},}\ }\href@noop {} {\bibfield
  {journal} {\bibinfo  {journal} {Journal of Materials Research}\ }\textbf
  {\bibinfo {volume} {31}},\ \bibinfo {pages} {832–844} (\bibinfo {year}
  {2016})}\BibitemShut {NoStop}%
\bibitem [{\citenamefont {Liu}\ \emph {et~al.}(2020)\citenamefont {Liu},
  \citenamefont {Li}, \citenamefont {Min}, \citenamefont {Chang}, \citenamefont
  {Shu}, \citenamefont {Ding},\ and\ \citenamefont {Yu}}]{Liu_2020}%
  \BibitemOpen
  \bibfield  {author} {\bibinfo {author} {\bibnamefont {Liu}, \bibfnamefont
  {P.}}, \bibinfo {author} {\bibnamefont {Li}, \bibfnamefont {X.}}, \bibinfo
  {author} {\bibnamefont {Min}, \bibfnamefont {P.}}, \bibinfo {author}
  {\bibnamefont {Chang}, \bibfnamefont {X.}}, \bibinfo {author} {\bibnamefont
  {Shu}, \bibfnamefont {C.}}, \bibinfo {author} {\bibnamefont {Ding},
  \bibfnamefont {Y.}}, and\ \bibinfo {author} {\bibnamefont {Yu}, \bibfnamefont
  {Z.-Z.}},\ }\bibfield  {title} {\enquote {\bibinfo {title} {3d
  lamellar-structured graphene aerogels for thermal interface composites with
  high through-plane thermal conductivity and fracture toughness},}\
  }\href@noop {} {\bibfield  {journal} {\bibinfo  {journal} {Nano-Micro
  Letters}\ }\textbf {\bibinfo {volume} {13}} (\bibinfo {year}
  {2020})}\BibitemShut {NoStop}%
\bibitem [{\citenamefont {Mahdizadeh}\ and\ \citenamefont
  {Akhlamadi}(2017)}]{chi_opt}%
  \BibitemOpen
  \bibfield  {author} {\bibinfo {author} {\bibnamefont {Mahdizadeh},
  \bibfnamefont {S.~J.}}and\ \bibinfo {author} {\bibnamefont {Akhlamadi},
  \bibfnamefont {G.}},\ }\bibfield  {title} {\enquote {\bibinfo {title}
  {Optimized tersoff empirical potential for germanene},}\ }\href@noop {}
  {\bibfield  {journal} {\bibinfo  {journal} {Journal of Molecular Graphics and
  Modelling}\ }\textbf {\bibinfo {volume} {72}},\ \bibinfo {pages} {1--5}
  (\bibinfo {year} {2017})}\BibitemShut {NoStop}%
\bibitem [{\citenamefont {{Mingjian Wen}}(2019{\natexlab{a}})}]{dunn1}%
  \BibitemOpen
  \bibfield  {author} {\bibinfo {author} {\bibnamefont {{Mingjian Wen}},},\
  }\href {https://doi.org/10.25950/44B7F4ED} {\enquote {\bibinfo {title}
  {Dropout uncertainty neural network (dunn) potential for condensed-matter
  carbon systems developed by wen and tadmor (2019) v000},}\ } (\bibinfo {year}
  {2019}{\natexlab{a}})\BibitemShut {NoStop}%
\bibitem [{\citenamefont {{Mingjian Wen}}(2019{\natexlab{b}})}]{dunn2}%
  \BibitemOpen
  \bibfield  {author} {\bibinfo {author} {\bibnamefont {{Mingjian Wen}},},\
  }\href {https://doi.org/10.25950/5CDB2C9F} {\enquote {\bibinfo {title}
  {Dropout uncertainty neural network (dunn) potential for condensed-matter
  carbon systems developed by wen and tadmor (2019) v000},}\ } (\bibinfo {year}
  {2019}{\natexlab{b}})\BibitemShut {NoStop}%
\bibitem [{\citenamefont {{Mingjian Wen}}(2019{\natexlab{c}})}]{dunn3}%
  \BibitemOpen
  \bibfield  {author} {\bibinfo {author} {\bibnamefont {{Mingjian Wen}},},\
  }\href {https://doi.org/10.25950/656F7A62} {\enquote {\bibinfo {title}
  {Dropout uncertainty neural network (dunn) potential for condensed-matter
  carbon systems developed by wen and tadmor (2019) v000},}\ } (\bibinfo {year}
  {2019}{\natexlab{c}})\BibitemShut {NoStop}%
\bibitem [{\citenamefont {Mortazavi}\ \emph {et~al.}(2022)\citenamefont
  {Mortazavi}, \citenamefont {Rajabpour}, \citenamefont {Zhuang}, \citenamefont
  {Rabczuk},\ and\ \citenamefont {Shapeev}}]{carbon_ml}%
  \BibitemOpen
  \bibfield  {author} {\bibinfo {author} {\bibnamefont {Mortazavi},
  \bibfnamefont {B.}}, \bibinfo {author} {\bibnamefont {Rajabpour},
  \bibfnamefont {A.}}, \bibinfo {author} {\bibnamefont {Zhuang}, \bibfnamefont
  {X.}}, \bibinfo {author} {\bibnamefont {Rabczuk}, \bibfnamefont {T.}}, and\
  \bibinfo {author} {\bibnamefont {Shapeev}, \bibfnamefont {A.~V.}},\
  }\bibfield  {title} {\enquote {\bibinfo {title} {Exploring thermal expansion
  of carbon-based nanosheets by machine-learning interatomic potentials},}\
  }\href {https://doi.org/10.1016/j.carbon.2021.10.059} {\bibfield  {journal}
  {\bibinfo  {journal} {Carbon}\ }\textbf {\bibinfo {volume} {186}},\ \bibinfo
  {pages} {501--508} (\bibinfo {year} {2022})}\BibitemShut {NoStop}%
\bibitem [{\citenamefont {Mortazavi}\ \emph {et~al.}(2021)\citenamefont
  {Mortazavi}, \citenamefont {Silani}, \citenamefont {Podryabinkin},
  \citenamefont {Rabczuk}, \citenamefont {Zhuang},\ and\ \citenamefont
  {Shapeev}}]{advmat}%
  \BibitemOpen
  \bibfield  {author} {\bibinfo {author} {\bibnamefont {Mortazavi},
  \bibfnamefont {B.}}, \bibinfo {author} {\bibnamefont {Silani}, \bibfnamefont
  {M.}}, \bibinfo {author} {\bibnamefont {Podryabinkin}, \bibfnamefont
  {E.~V.}}, \bibinfo {author} {\bibnamefont {Rabczuk}, \bibfnamefont {T.}},
  \bibinfo {author} {\bibnamefont {Zhuang}, \bibfnamefont {X.}}, and\ \bibinfo
  {author} {\bibnamefont {Shapeev}, \bibfnamefont {A.~V.}},\ }\bibfield
  {title} {\enquote {\bibinfo {title} {First-principles multiscale modeling of
  mechanical properties in graphene/borophene heterostructures empowered by
  machine-learning interatomic potentials},}\ }\href
  {https://doi.org/10.1002/adma.202102807} {\bibfield  {journal} {\bibinfo
  {journal} {Advanced Materials}\ }\textbf {\bibinfo {volume} {33}},\ \bibinfo
  {pages} {2102807} (\bibinfo {year} {2021})}\BibitemShut {NoStop}%
\bibitem [{\citenamefont {Ng}, \citenamefont {Yeo},\ and\ \citenamefont
  {Liu}(2012)}]{jj}%
  \BibitemOpen
  \bibfield  {author} {\bibinfo {author} {\bibnamefont {Ng}, \bibfnamefont
  {T.}}, \bibinfo {author} {\bibnamefont {Yeo}, \bibfnamefont {J.}}, and\
  \bibinfo {author} {\bibnamefont {Liu}, \bibfnamefont {Z.}},\ }\bibfield
  {title} {\enquote {\bibinfo {title} {A molecular dynamics study of the
  thermal conductivity of graphene nanoribbons containing dispersed
  stone{\textendash}thrower{\textendash}wales defects},}\ }\href@noop {}
  {\bibfield  {journal} {\bibinfo  {journal} {Carbon}\ }\textbf {\bibinfo
  {volume} {50}},\ \bibinfo {pages} {4887--4893} (\bibinfo {year}
  {2012})}\BibitemShut {NoStop}%
\bibitem [{\citenamefont {Novikov}\ \emph {et~al.}(2022)\citenamefont
  {Novikov}, \citenamefont {Grabowski}, \citenamefont {Körmann},\ and\
  \citenamefont {Shapeev}}]{mlip_mag_vibration}%
  \BibitemOpen
  \bibfield  {author} {\bibinfo {author} {\bibnamefont {Novikov}, \bibfnamefont
  {I.}}, \bibinfo {author} {\bibnamefont {Grabowski}, \bibfnamefont {B.}},
  \bibinfo {author} {\bibnamefont {Körmann}, \bibfnamefont {F.}}, and\
  \bibinfo {author} {\bibnamefont {Shapeev}, \bibfnamefont {A.}},\ }\bibfield
  {title} {\enquote {\bibinfo {title} {Magnetic moment tensor potentials for
  collinear spin-polarized materials reproduce different magnetic states of bcc
  fe},}\ }\href {https://doi.org/10.1038/s41524-022-00696-9} {\bibfield
  {journal} {\bibinfo  {journal} {npj Computational Materials}\ }\textbf
  {\bibinfo {volume} {8}} (\bibinfo {year} {2022}),\
  10.1038/s41524-022-00696-9}\BibitemShut {NoStop}%
\bibitem [{\citenamefont {Novikov}\ \emph {et~al.}(2021)\citenamefont
  {Novikov}, \citenamefont {Gubaev}, \citenamefont {Podryabinkin},\ and\
  \citenamefont {Shapeev}}]{mlip}%
  \BibitemOpen
  \bibfield  {author} {\bibinfo {author} {\bibnamefont {Novikov}, \bibfnamefont
  {I.~S.}}, \bibinfo {author} {\bibnamefont {Gubaev}, \bibfnamefont {K.}},
  \bibinfo {author} {\bibnamefont {Podryabinkin}, \bibfnamefont {E.~V.}}, and\
  \bibinfo {author} {\bibnamefont {Shapeev}, \bibfnamefont {A.~V.}},\
  }\bibfield  {title} {\enquote {\bibinfo {title} {The {MLIP} package: moment
  tensor potentials with {MPI} and active learning},}\ }\href
  {https://doi.org/10.1088/2632-2153/abc9fe} {\bibfield  {journal} {\bibinfo
  {journal} {Machine Learning: Science and Technology}\ }\textbf {\bibinfo
  {volume} {2}},\ \bibinfo {pages} {025002} (\bibinfo {year}
  {2021})}\BibitemShut {NoStop}%
\bibitem [{\citenamefont {Novoselov}\ \emph {et~al.}(2004)\citenamefont
  {Novoselov}, \citenamefont {Geim}, \citenamefont {Morozov}, \citenamefont
  {Jiang}, \citenamefont {Zhang}, \citenamefont {Dubonos}, \citenamefont
  {Grigorieva},\ and\ \citenamefont {Firsov}}]{first_graphene_paper}%
  \BibitemOpen
  \bibfield  {author} {\bibinfo {author} {\bibnamefont {Novoselov},
  \bibfnamefont {K.~S.}}, \bibinfo {author} {\bibnamefont {Geim}, \bibfnamefont
  {A.~K.}}, \bibinfo {author} {\bibnamefont {Morozov}, \bibfnamefont {S.~V.}},
  \bibinfo {author} {\bibnamefont {Jiang}, \bibfnamefont {D.}}, \bibinfo
  {author} {\bibnamefont {Zhang}, \bibfnamefont {Y.}}, \bibinfo {author}
  {\bibnamefont {Dubonos}, \bibfnamefont {S.~V.}}, \bibinfo {author}
  {\bibnamefont {Grigorieva}, \bibfnamefont {I.~V.}}, and\ \bibinfo {author}
  {\bibnamefont {Firsov}, \bibfnamefont {A.~A.}},\ }\bibfield  {title}
  {\enquote {\bibinfo {title} {Electric field effect in atomically thin carbon
  films},}\ }\href@noop {} {\bibfield  {journal} {\bibinfo  {journal}
  {Science}\ }\textbf {\bibinfo {volume} {306}},\ \bibinfo {pages} {666--669}
  (\bibinfo {year} {2004})}\BibitemShut {NoStop}%
\bibitem [{\citenamefont {Obeng}\ and\ \citenamefont
  {Srinivasan}(2011)}]{semiconductor}%
  \BibitemOpen
  \bibfield  {author} {\bibinfo {author} {\bibnamefont {Obeng}, \bibfnamefont
  {Y.}}and\ \bibinfo {author} {\bibnamefont {Srinivasan}, \bibfnamefont {P.}},\
  }\bibfield  {title} {\enquote {\bibinfo {title} {Graphene: Is it the future
  for semiconductors? an overview of the material, devices, and
  applications},}\ }\href@noop {} {\bibfield  {journal} {\bibinfo  {journal}
  {Interface magazine}\ }\textbf {\bibinfo {volume} {20}},\ \bibinfo {pages}
  {47--52} (\bibinfo {year} {2011})}\BibitemShut {NoStop}%
\bibitem [{\citenamefont {O'Connor}, \citenamefont {Andzelm},\ and\
  \citenamefont {Robbins}(2015)}]{airebo-m}%
  \BibitemOpen
  \bibfield  {author} {\bibinfo {author} {\bibnamefont {O'Connor},
  \bibfnamefont {T.~C.}}, \bibinfo {author} {\bibnamefont {Andzelm},
  \bibfnamefont {J.}}, and\ \bibinfo {author} {\bibnamefont {Robbins},
  \bibfnamefont {M.~O.}},\ }\bibfield  {title} {\enquote {\bibinfo {title}
  {{AIREBO}-m: A reactive model for hydrocarbons at extreme pressures},}\
  }\href {https://doi.org/10.1063/1.4905549} {\bibfield  {journal} {\bibinfo
  {journal} {The Journal of Chemical Physics}\ }\textbf {\bibinfo {volume}
  {142}},\ \bibinfo {pages} {024903} (\bibinfo {year} {2015})}\BibitemShut
  {NoStop}%
\bibitem [{\citenamefont {Papageorgiou}, \citenamefont {Kinloch},\ and\
  \citenamefont {Young}(2017)}]{strength1}%
  \BibitemOpen
  \bibfield  {author} {\bibinfo {author} {\bibnamefont {Papageorgiou},
  \bibfnamefont {D.~G.}}, \bibinfo {author} {\bibnamefont {Kinloch},
  \bibfnamefont {I.~A.}}, and\ \bibinfo {author} {\bibnamefont {Young},
  \bibfnamefont {R.~J.}},\ }\bibfield  {title} {\enquote {\bibinfo {title}
  {Mechanical properties of graphene and graphene-based nanocomposites},}\
  }\href@noop {} {\bibfield  {journal} {\bibinfo  {journal} {Progress in
  Materials Science}\ }\textbf {\bibinfo {volume} {90}},\ \bibinfo {pages}
  {75--127} (\bibinfo {year} {2017})}\BibitemShut {NoStop}%
\bibitem [{\citenamefont {Plimpton}(1995)}]{lammps}%
  \BibitemOpen
  \bibfield  {author} {\bibinfo {author} {\bibnamefont {Plimpton},
  \bibfnamefont {S.}},\ }\bibfield  {title} {\enquote {\bibinfo {title} {Fast
  parallel algorithms for short-range molecular dynamics},}\ }\href@noop {}
  {\bibfield  {journal} {\bibinfo  {journal} {Journal of computational
  physics}\ }\textbf {\bibinfo {volume} {117}},\ \bibinfo {pages} {1--19}
  (\bibinfo {year} {1995})}\BibitemShut {NoStop}%
\bibitem [{\citenamefont {Pugno}\ \emph {et~al.}(2008)\citenamefont {Pugno},
  \citenamefont {Carpinteri}, \citenamefont {Ippolito}, \citenamefont
  {Mattoni},\ and\ \citenamefont {Colombo}}]{qfm_md}%
  \BibitemOpen
  \bibfield  {author} {\bibinfo {author} {\bibnamefont {Pugno}, \bibfnamefont
  {N.}}, \bibinfo {author} {\bibnamefont {Carpinteri}, \bibfnamefont {A.}},
  \bibinfo {author} {\bibnamefont {Ippolito}, \bibfnamefont {M.}}, \bibinfo
  {author} {\bibnamefont {Mattoni}, \bibfnamefont {A.}}, and\ \bibinfo {author}
  {\bibnamefont {Colombo}, \bibfnamefont {L.}},\ }\bibfield  {title} {\enquote
  {\bibinfo {title} {Atomistic fracture: {QFM} vs. {MD}},}\ }\href
  {https://doi.org/10.1016/j.engfracmech.2007.01.028} {\bibfield  {journal}
  {\bibinfo  {journal} {Engineering Fracture Mechanics}\ }\textbf {\bibinfo
  {volume} {75}},\ \bibinfo {pages} {1794--1803} (\bibinfo {year}
  {2008})}\BibitemShut {NoStop}%
\bibitem [{\citenamefont {Pugno}\ and\ \citenamefont {Ruoff}(2004)}]{qfm}%
  \BibitemOpen
  \bibfield  {author} {\bibinfo {author} {\bibnamefont {Pugno}, \bibfnamefont
  {N.~M.}}and\ \bibinfo {author} {\bibnamefont {Ruoff}, \bibfnamefont
  {R.~S.}},\ }\bibfield  {title} {\enquote {\bibinfo {title} {Quantized
  fracture mechanics},}\ }\href {https://doi.org/10.1080/14786430412331280382}
  {\bibfield  {journal} {\bibinfo  {journal} {Philosophical Magazine}\ }\textbf
  {\bibinfo {volume} {84}},\ \bibinfo {pages} {2829--2845} (\bibinfo {year}
  {2004})}\BibitemShut {NoStop}%
\bibitem [{\citenamefont {Rose}, \citenamefont {Smith},\ and\ \citenamefont
  {Ferrante}(1983)}]{rose1983}%
  \BibitemOpen
  \bibfield  {author} {\bibinfo {author} {\bibnamefont {Rose}, \bibfnamefont
  {J.~H.}}, \bibinfo {author} {\bibnamefont {Smith}, \bibfnamefont {J.~R.}},
  and\ \bibinfo {author} {\bibnamefont {Ferrante}, \bibfnamefont {J.}},\
  }\bibfield  {title} {\enquote {\bibinfo {title} {Universal features of
  bonding in metals},}\ }\href@noop {} {\bibfield  {journal} {\bibinfo
  {journal} {Phys. Rev. B}\ }\textbf {\bibinfo {volume} {28}},\ \bibinfo
  {pages} {1835--1845} (\bibinfo {year} {1983})}\BibitemShut {NoStop}%
\bibitem [{\citenamefont {Saumya}\ \emph {et~al.}(2020)\citenamefont {Saumya},
  \citenamefont {Gupta}, \citenamefont {Roy},\ and\ \citenamefont
  {Dey}}]{iop_conference}%
  \BibitemOpen
  \bibfield  {author} {\bibinfo {author} {\bibnamefont {Saumya}, \bibfnamefont
  {K.}}, \bibinfo {author} {\bibnamefont {Gupta}, \bibfnamefont {K.~K.}},
  \bibinfo {author} {\bibnamefont {Roy}, \bibfnamefont {A.}}, and\ \bibinfo
  {author} {\bibnamefont {Dey}, \bibfnamefont {S.}},\ }\bibfield  {title}
  {\enquote {\bibinfo {title} {Effect of spatial distribution of nanopores on
  mechanical properties of mono layer graphene},}\ }\href
  {https://doi.org/10.1088/1757-899x/872/1/012187} {\bibfield  {journal}
  {\bibinfo  {journal} {{IOP} Conference Series: Materials Science and
  Engineering}\ }\textbf {\bibinfo {volume} {872}},\ \bibinfo {pages} {012187}
  (\bibinfo {year} {2020})}\BibitemShut {NoStop}%
\bibitem [{\citenamefont {Shareena}\ \emph {et~al.}(2018)\citenamefont
  {Shareena}, \citenamefont {McShan}, \citenamefont {Dasmahapatra},\ and\
  \citenamefont {Tchounwou}}]{biomed_2}%
  \BibitemOpen
  \bibfield  {author} {\bibinfo {author} {\bibnamefont {Shareena},
  \bibfnamefont {T.~P.~D.}}, \bibinfo {author} {\bibnamefont {McShan},
  \bibfnamefont {D.}}, \bibinfo {author} {\bibnamefont {Dasmahapatra},
  \bibfnamefont {A.~K.}}, and\ \bibinfo {author} {\bibnamefont {Tchounwou},
  \bibfnamefont {P.~B.}},\ }\bibfield  {title} {\enquote {\bibinfo {title} {A
  review on graphene-based nanomaterials in biomedical applications and risks
  in environment and health},}\ }\href@noop {} {\bibfield  {journal} {\bibinfo
  {journal} {Nano-Micro Letters}\ }\textbf {\bibinfo {volume} {10}} (\bibinfo
  {year} {2018})}\BibitemShut {NoStop}%
\bibitem [{\citenamefont {Si}\ \emph {et~al.}(2017)\citenamefont {Si},
  \citenamefont {Wang}, \citenamefont {Fan}, \citenamefont {Feng},\ and\
  \citenamefont {Cao}}]{potential_model}%
  \BibitemOpen
  \bibfield  {author} {\bibinfo {author} {\bibnamefont {Si}, \bibfnamefont
  {C.}}, \bibinfo {author} {\bibnamefont {Wang}, \bibfnamefont {X.-D.}},
  \bibinfo {author} {\bibnamefont {Fan}, \bibfnamefont {Z.}}, \bibinfo {author}
  {\bibnamefont {Feng}, \bibfnamefont {Z.-H.}}, and\ \bibinfo {author}
  {\bibnamefont {Cao}, \bibfnamefont {B.-Y.}},\ }\bibfield  {title} {\enquote
  {\bibinfo {title} {Impacts of potential models on calculating the thermal
  conductivity of graphene using non-equilibrium molecular dynamics
  simulations},}\ }\href@noop {} {\bibfield  {journal} {\bibinfo  {journal}
  {International Journal of Heat and Mass Transfer}\ }\textbf {\bibinfo
  {volume} {107}},\ \bibinfo {pages} {450--460} (\bibinfo {year}
  {2017})}\BibitemShut {NoStop}%
\bibitem [{\citenamefont {Song}\ \emph {et~al.}(2018)\citenamefont {Song},
  \citenamefont {Liu}, \citenamefont {Liu}, \citenamefont {Wu}, \citenamefont
  {Cheng},\ and\ \citenamefont {Kang}}]{2d_thermal}%
  \BibitemOpen
  \bibfield  {author} {\bibinfo {author} {\bibnamefont {Song}, \bibfnamefont
  {H.}}, \bibinfo {author} {\bibnamefont {Liu}, \bibfnamefont {J.}}, \bibinfo
  {author} {\bibnamefont {Liu}, \bibfnamefont {B.}}, \bibinfo {author}
  {\bibnamefont {Wu}, \bibfnamefont {J.}}, \bibinfo {author} {\bibnamefont
  {Cheng}, \bibfnamefont {H.-M.}}, and\ \bibinfo {author} {\bibnamefont {Kang},
  \bibfnamefont {F.}},\ }\bibfield  {title} {\enquote {\bibinfo {title}
  {Two-dimensional materials for thermal management applications},}\
  }\href@noop {} {\bibfield  {journal} {\bibinfo  {journal} {Joule}\ }\textbf
  {\bibinfo {volume} {2}},\ \bibinfo {pages} {442--463} (\bibinfo {year}
  {2018})}\BibitemShut {NoStop}%
\bibitem [{\citenamefont {Stuart}, \citenamefont {Tutein},\ and\ \citenamefont
  {Harrison}(2000)}]{airebo}%
  \BibitemOpen
  \bibfield  {author} {\bibinfo {author} {\bibnamefont {Stuart}, \bibfnamefont
  {S.~J.}}, \bibinfo {author} {\bibnamefont {Tutein}, \bibfnamefont {A.~B.}},
  and\ \bibinfo {author} {\bibnamefont {Harrison}, \bibfnamefont {J.~A.}},\
  }\bibfield  {title} {\enquote {\bibinfo {title} {A reactive potential for
  hydrocarbons with intermolecular interactions},}\ }\href
  {https://doi.org/10.1063/1.481208} {\bibfield  {journal} {\bibinfo  {journal}
  {The Journal of Chemical Physics}\ }\textbf {\bibinfo {volume} {112}},\
  \bibinfo {pages} {6472--6486} (\bibinfo {year} {2000})}\BibitemShut {NoStop}%
\bibitem [{\citenamefont {Tersoff}(1986)}]{tersoff1986}%
  \BibitemOpen
  \bibfield  {author} {\bibinfo {author} {\bibnamefont {Tersoff}, \bibfnamefont
  {J.}},\ }\bibfield  {title} {\enquote {\bibinfo {title} {New empirical model
  for the structural properties of silicon},}\ }\href@noop {} {\bibfield
  {journal} {\bibinfo  {journal} {Phys. Rev. Lett.}\ }\textbf {\bibinfo
  {volume} {56}},\ \bibinfo {pages} {632--635} (\bibinfo {year}
  {1986})}\BibitemShut {NoStop}%
\bibitem [{\citenamefont {Tersoff}(1988)}]{Tersoff_1988}%
  \BibitemOpen
  \bibfield  {author} {\bibinfo {author} {\bibnamefont {Tersoff}, \bibfnamefont
  {J.}},\ }\bibfield  {title} {\enquote {\bibinfo {title} {New empirical
  approach for the structure and energy of covalent systems},}\ }\href@noop {}
  {\bibfield  {journal} {\bibinfo  {journal} {Physical Review B}\ }\textbf
  {\bibinfo {volume} {37}},\ \bibinfo {pages} {6991--7000} (\bibinfo {year}
  {1988})}\BibitemShut {NoStop}%
\bibitem [{\citenamefont {Wan}\ \emph {et~al.}(2021)\citenamefont {Wan},
  \citenamefont {Chen}, \citenamefont {Fang}, \citenamefont {Wang},
  \citenamefont {Xu}, \citenamefont {Jiang}, \citenamefont {Baughman},\ and\
  \citenamefont {Cheng}}]{strength2}%
  \BibitemOpen
  \bibfield  {author} {\bibinfo {author} {\bibnamefont {Wan}, \bibfnamefont
  {S.}}, \bibinfo {author} {\bibnamefont {Chen}, \bibfnamefont {Y.}}, \bibinfo
  {author} {\bibnamefont {Fang}, \bibfnamefont {S.}}, \bibinfo {author}
  {\bibnamefont {Wang}, \bibfnamefont {S.}}, \bibinfo {author} {\bibnamefont
  {Xu}, \bibfnamefont {Z.}}, \bibinfo {author} {\bibnamefont {Jiang},
  \bibfnamefont {L.}}, \bibinfo {author} {\bibnamefont {Baughman},
  \bibfnamefont {R.~H.}}, and\ \bibinfo {author} {\bibnamefont {Cheng},
  \bibfnamefont {Q.}},\ }\bibfield  {title} {\enquote {\bibinfo {title}
  {High-strength scalable graphene sheets by freezing stretch-induced
  alignment},}\ }\href@noop {} {\bibfield  {journal} {\bibinfo  {journal}
  {Nature Materials}\ }\textbf {\bibinfo {volume} {20}},\ \bibinfo {pages}
  {624--631} (\bibinfo {year} {2021})}\BibitemShut {NoStop}%
\bibitem [{\citenamefont {Wang}\ \emph {et~al.}(2009)\citenamefont {Wang},
  \citenamefont {Li}, \citenamefont {Too},\ and\ \citenamefont
  {Wallace}}]{graphene_battery_1}%
  \BibitemOpen
  \bibfield  {author} {\bibinfo {author} {\bibnamefont {Wang}, \bibfnamefont
  {C.}}, \bibinfo {author} {\bibnamefont {Li}, \bibfnamefont {D.}}, \bibinfo
  {author} {\bibnamefont {Too}, \bibfnamefont {C.~O.}}, and\ \bibinfo {author}
  {\bibnamefont {Wallace}, \bibfnamefont {G.~G.}},\ }\bibfield  {title}
  {\enquote {\bibinfo {title} {Electrochemical properties of graphene paper
  electrodes used in lithium batteries},}\ }\href@noop {} {\bibfield  {journal}
  {\bibinfo  {journal} {Chemistry of Materials}\ }\textbf {\bibinfo {volume}
  {21}},\ \bibinfo {pages} {2604--2606} (\bibinfo {year} {2009})}\BibitemShut
  {NoStop}%
\bibitem [{\citenamefont {Wang}\ \emph {et~al.}(2018)\citenamefont {Wang},
  \citenamefont {Zhang}, \citenamefont {Han},\ and\ \citenamefont
  {E}}]{deepmd}%
  \BibitemOpen
  \bibfield  {author} {\bibinfo {author} {\bibnamefont {Wang}, \bibfnamefont
  {H.}}, \bibinfo {author} {\bibnamefont {Zhang}, \bibfnamefont {L.}}, \bibinfo
  {author} {\bibnamefont {Han}, \bibfnamefont {J.}}, and\ \bibinfo {author}
  {\bibnamefont {E}, \bibfnamefont {W.}},\ }\bibfield  {title} {\enquote
  {\bibinfo {title} {{DeePMD}-kit: A deep learning package for many-body
  potential energy representation and molecular dynamics},}\ }\href
  {https://doi.org/10.1016/j.cpc.2018.03.016} {\bibfield  {journal} {\bibinfo
  {journal} {Computer Physics Communications}\ }\textbf {\bibinfo {volume}
  {228}},\ \bibinfo {pages} {178--184} (\bibinfo {year} {2018})}\BibitemShut
  {NoStop}%
\bibitem [{\citenamefont {Wei}\ and\ \citenamefont {Kysar}(2012)}]{elastic1}%
  \BibitemOpen
  \bibfield  {author} {\bibinfo {author} {\bibnamefont {Wei}, \bibfnamefont
  {X.}}and\ \bibinfo {author} {\bibnamefont {Kysar}, \bibfnamefont {J.~W.}},\
  }\bibfield  {title} {\enquote {\bibinfo {title} {Experimental validation of
  multiscale modeling of indentation of suspended circular graphene
  membranes},}\ }\href@noop {} {\bibfield  {journal} {\bibinfo  {journal}
  {International Journal of Solids and Structures}\ }\textbf {\bibinfo {volume}
  {49}},\ \bibinfo {pages} {3201--3209} (\bibinfo {year} {2012})}\BibitemShut
  {NoStop}%
\bibitem [{\citenamefont {Wei}\ \emph {et~al.}(2015)\citenamefont {Wei},
  \citenamefont {Xiao}, \citenamefont {Li}, \citenamefont {Tang}, \citenamefont
  {Chen}, \citenamefont {Bando},\ and\ \citenamefont
  {Golberg}}]{fracture_toughness_12}%
  \BibitemOpen
  \bibfield  {author} {\bibinfo {author} {\bibnamefont {Wei}, \bibfnamefont
  {X.}}, \bibinfo {author} {\bibnamefont {Xiao}, \bibfnamefont {S.}}, \bibinfo
  {author} {\bibnamefont {Li}, \bibfnamefont {F.}}, \bibinfo {author}
  {\bibnamefont {Tang}, \bibfnamefont {D.-M.}}, \bibinfo {author} {\bibnamefont
  {Chen}, \bibfnamefont {Q.}}, \bibinfo {author} {\bibnamefont {Bando},
  \bibfnamefont {Y.}}, and\ \bibinfo {author} {\bibnamefont {Golberg},
  \bibfnamefont {D.}},\ }\bibfield  {title} {\enquote {\bibinfo {title}
  {Comparative fracture toughness of multilayer graphenes and boronitrenes},}\
  }\href {https://doi.org/10.1021/nl5042066} {\bibfield  {journal} {\bibinfo
  {journal} {Nano Letters}\ }\textbf {\bibinfo {volume} {15}},\ \bibinfo
  {pages} {689--694} (\bibinfo {year} {2015})}\BibitemShut {NoStop}%
\bibitem [{\citenamefont {Wei}\ \emph {et~al.}(2012)\citenamefont {Wei},
  \citenamefont {Wu}, \citenamefont {Yin}, \citenamefont {Shi}, \citenamefont
  {Yang},\ and\ \citenamefont {Dresselhaus}}]{defect_mech2}%
  \BibitemOpen
  \bibfield  {author} {\bibinfo {author} {\bibnamefont {Wei}, \bibfnamefont
  {Y.}}, \bibinfo {author} {\bibnamefont {Wu}, \bibfnamefont {J.}}, \bibinfo
  {author} {\bibnamefont {Yin}, \bibfnamefont {H.}}, \bibinfo {author}
  {\bibnamefont {Shi}, \bibfnamefont {X.}}, \bibinfo {author} {\bibnamefont
  {Yang}, \bibfnamefont {R.}}, and\ \bibinfo {author} {\bibnamefont
  {Dresselhaus}, \bibfnamefont {M.}},\ }\bibfield  {title} {\enquote {\bibinfo
  {title} {The nature of strength enhancement and weakening by
  pentagon{\textendash}heptagon defects in~graphene},}\ }\href@noop {}
  {\bibfield  {journal} {\bibinfo  {journal} {Nature Materials}\ }\textbf
  {\bibinfo {volume} {11}},\ \bibinfo {pages} {759--763} (\bibinfo {year}
  {2012})}\BibitemShut {NoStop}%
\bibitem [{\citenamefont {Wei}\ and\ \citenamefont {Yang}(2018)}]{nsr_review}%
  \BibitemOpen
  \bibfield  {author} {\bibinfo {author} {\bibnamefont {Wei}, \bibfnamefont
  {Y.}}and\ \bibinfo {author} {\bibnamefont {Yang}, \bibfnamefont {R.}},\
  }\bibfield  {title} {\enquote {\bibinfo {title} {Nanomechanics of
  graphene},}\ }\href@noop {} {\bibfield  {journal} {\bibinfo  {journal}
  {National Science Review}\ }\textbf {\bibinfo {volume} {6}},\ \bibinfo
  {pages} {324--348} (\bibinfo {year} {2018})}\BibitemShut {NoStop}%
\bibitem [{\citenamefont {Wen}\ and\ \citenamefont
  {Tadmor}(2019)}]{dunn_graphene}%
  \BibitemOpen
  \bibfield  {author} {\bibinfo {author} {\bibnamefont {Wen}, \bibfnamefont
  {M.}}and\ \bibinfo {author} {\bibnamefont {Tadmor}, \bibfnamefont {E.~B.}},\
  }\bibfield  {title} {\enquote {\bibinfo {title} {Hybrid neural network
  potential for multilayer graphene},}\ }\href
  {https://doi.org/10.1103/physrevb.100.195419} {\bibfield  {journal} {\bibinfo
   {journal} {Physical Review B}\ }\textbf {\bibinfo {volume} {100}} (\bibinfo
  {year} {2019}),\ 10.1103/physrevb.100.195419}\BibitemShut {NoStop}%
\bibitem [{\citenamefont {Wen}\ and\ \citenamefont {Tadmor}(2020)}]{dunn}%
  \BibitemOpen
  \bibfield  {author} {\bibinfo {author} {\bibnamefont {Wen}, \bibfnamefont
  {M.}}and\ \bibinfo {author} {\bibnamefont {Tadmor}, \bibfnamefont {E.~B.}},\
  }\bibfield  {title} {\enquote {\bibinfo {title} {Uncertainty quantification
  in molecular simulations with dropout neural network potentials},}\ }\href
  {https://doi.org/10.1038/s41524-020-00390-8} {\bibfield  {journal} {\bibinfo
  {journal} {npj Computational Materials}\ }\textbf {\bibinfo {volume} {6}}
  (\bibinfo {year} {2020}),\ 10.1038/s41524-020-00390-8}\BibitemShut {NoStop}%
\bibitem [{\citenamefont {Xie}\ \emph {et~al.}(2018)\citenamefont {Xie},
  \citenamefont {Wang}, \citenamefont {Zhang}, \citenamefont {Wang},\ and\
  \citenamefont {Luo}}]{semiconductor2}%
  \BibitemOpen
  \bibfield  {author} {\bibinfo {author} {\bibnamefont {Xie}, \bibfnamefont
  {C.}}, \bibinfo {author} {\bibnamefont {Wang}, \bibfnamefont {Y.}}, \bibinfo
  {author} {\bibnamefont {Zhang}, \bibfnamefont {Z.-X.}}, \bibinfo {author}
  {\bibnamefont {Wang}, \bibfnamefont {D.}}, and\ \bibinfo {author}
  {\bibnamefont {Luo}, \bibfnamefont {L.-B.}},\ }\bibfield  {title} {\enquote
  {\bibinfo {title} {Graphene/semiconductor hybrid heterostructures for
  optoelectronic device applications},}\ }\href@noop {} {\bibfield  {journal}
  {\bibinfo  {journal} {Nano Today}\ }\textbf {\bibinfo {volume} {19}},\
  \bibinfo {pages} {41--83} (\bibinfo {year} {2018})}\BibitemShut {NoStop}%
\bibitem [{\citenamefont {Xu}\ \emph {et~al.}(2012)\citenamefont {Xu},
  \citenamefont {Wei}, \citenamefont {Zheng}, \citenamefont {Fan},
  \citenamefont {Wang},\ and\ \citenamefont {Zheng}}]{Thermal_Mechanical}%
  \BibitemOpen
  \bibfield  {author} {\bibinfo {author} {\bibnamefont {Xu}, \bibfnamefont
  {L.}}, \bibinfo {author} {\bibnamefont {Wei}, \bibfnamefont {N.}}, \bibinfo
  {author} {\bibnamefont {Zheng}, \bibfnamefont {Y.}}, \bibinfo {author}
  {\bibnamefont {Fan}, \bibfnamefont {Z.}}, \bibinfo {author} {\bibnamefont
  {Wang}, \bibfnamefont {H.-Q.}}, and\ \bibinfo {author} {\bibnamefont {Zheng},
  \bibfnamefont {J.-C.}},\ }\bibfield  {title} {\enquote {\bibinfo {title}
  {Graphene-nanotube 3d networks: intriguing thermal and mechanical
  properties},}\ }\href@noop {} {\bibfield  {journal} {\bibinfo  {journal} {J.
  Mater. Chem.}\ }\textbf {\bibinfo {volume} {22}},\ \bibinfo {pages}
  {1435--1444} (\bibinfo {year} {2012})}\BibitemShut {NoStop}%
\bibitem [{\citenamefont {Xu}\ and\ \citenamefont {Buehler}(2009)}]{mech1}%
  \BibitemOpen
  \bibfield  {author} {\bibinfo {author} {\bibnamefont {Xu}, \bibfnamefont
  {Z.}}and\ \bibinfo {author} {\bibnamefont {Buehler}, \bibfnamefont {M.~J.}},\
  }\bibfield  {title} {\enquote {\bibinfo {title} {Strain controlled
  thermomutability of single-walled carbon nanotubes},}\ }\href@noop {}
  {\bibfield  {journal} {\bibinfo  {journal} {Nanotechnology}\ }\textbf
  {\bibinfo {volume} {20}},\ \bibinfo {pages} {185701} (\bibinfo {year}
  {2009})}\BibitemShut {NoStop}%
\bibitem [{\citenamefont {Yang}\ \emph {et~al.}(2013)\citenamefont {Yang},
  \citenamefont {Asiri}, \citenamefont {Tang}, \citenamefont {Du},\ and\
  \citenamefont {Lin}}]{biomed_1}%
  \BibitemOpen
  \bibfield  {author} {\bibinfo {author} {\bibnamefont {Yang}, \bibfnamefont
  {Y.}}, \bibinfo {author} {\bibnamefont {Asiri}, \bibfnamefont {A.~M.}},
  \bibinfo {author} {\bibnamefont {Tang}, \bibfnamefont {Z.}}, \bibinfo
  {author} {\bibnamefont {Du}, \bibfnamefont {D.}}, and\ \bibinfo {author}
  {\bibnamefont {Lin}, \bibfnamefont {Y.}},\ }\bibfield  {title} {\enquote
  {\bibinfo {title} {Graphene based materials for biomedical applications},}\
  }\href@noop {} {\bibfield  {journal} {\bibinfo  {journal} {Materials Today}\
  }\textbf {\bibinfo {volume} {16}},\ \bibinfo {pages} {365--373} (\bibinfo
  {year} {2013})}\BibitemShut {NoStop}%
\bibitem [{\citenamefont {Yanovsky}\ \emph {et~al.}(2009)\citenamefont
  {Yanovsky}, \citenamefont {Nikitina}, \citenamefont {Karnet},\ and\
  \citenamefont {Nikitin}}]{fracture_quantum}%
  \BibitemOpen
  \bibfield  {author} {\bibinfo {author} {\bibnamefont {Yanovsky},
  \bibfnamefont {Y.}}, \bibinfo {author} {\bibnamefont {Nikitina},
  \bibfnamefont {E.}}, \bibinfo {author} {\bibnamefont {Karnet}, \bibfnamefont
  {Y.}}, and\ \bibinfo {author} {\bibnamefont {Nikitin}, \bibfnamefont {S.}},\
  }\bibfield  {title} {\enquote {\bibinfo {title} {Quantum mechanics study of
  the mechanism of deformation and fracture of graphene},}\ }\href@noop {}
  {\bibfield  {journal} {\bibinfo  {journal} {Physical Mesomechanics}\ }\textbf
  {\bibinfo {volume} {12}},\ \bibinfo {pages} {254--262} (\bibinfo {year}
  {2009})}\BibitemShut {NoStop}%
\bibitem [{\citenamefont {Yoo}, \citenamefont {Xu},\ and\ \citenamefont
  {Ding}(2021)}]{heat_treat}%
  \BibitemOpen
  \bibfield  {author} {\bibinfo {author} {\bibnamefont {Yoo}, \bibfnamefont
  {B.}}, \bibinfo {author} {\bibnamefont {Xu}, \bibfnamefont {Z.}}, and\
  \bibinfo {author} {\bibnamefont {Ding}, \bibfnamefont {F.}},\ }\bibfield
  {title} {\enquote {\bibinfo {title} {How single-walled carbon nanotubes are
  transformed into multiwalled carbon nanotubes during heat treatment},}\
  }\href@noop {} {\bibfield  {journal} {\bibinfo  {journal} {{ACS} Omega}\
  }\textbf {\bibinfo {volume} {6}},\ \bibinfo {pages} {4074--4079} (\bibinfo
  {year} {2021})}\BibitemShut {NoStop}%
\bibitem [{\citenamefont {Yuan}\ and\ \citenamefont
  {Kalkhof}(2000)}]{thermal_couple_fracture}%
  \BibitemOpen
  \bibfield  {author} {\bibinfo {author} {\bibnamefont {Yuan}, \bibfnamefont
  {H.}}and\ \bibinfo {author} {\bibnamefont {Kalkhof}, \bibfnamefont {D.}},\
  }\bibfield  {title} {\enquote {\bibinfo {title} {Effects of temperature
  gradients on crack characterisation under thermal-mechanical loading
  conditions},}\ }\href@noop {} {\bibfield  {journal} {\bibinfo  {journal}
  {International Journal of Fracture}\ }\textbf {\bibinfo {volume} {100}},\
  \bibinfo {pages} {355--377} (\bibinfo {year} {2000})}\BibitemShut {NoStop}%
\bibitem [{\citenamefont {Zhang}, \citenamefont {Zhao},\ and\ \citenamefont
  {Lu}(2012)}]{defect_mech3}%
  \BibitemOpen
  \bibfield  {author} {\bibinfo {author} {\bibnamefont {Zhang}, \bibfnamefont
  {J.}}, \bibinfo {author} {\bibnamefont {Zhao}, \bibfnamefont {J.}}, and\
  \bibinfo {author} {\bibnamefont {Lu}, \bibfnamefont {J.}},\ }\bibfield
  {title} {\enquote {\bibinfo {title} {Intrinsic strength and failure behaviors
  of graphene grain boundaries},}\ }\href@noop {} {\bibfield  {journal}
  {\bibinfo  {journal} {{ACS} Nano}\ }\textbf {\bibinfo {volume} {6}},\
  \bibinfo {pages} {2704--2711} (\bibinfo {year} {2012})}\BibitemShut {NoStop}%
\bibitem [{\citenamefont {Zhang}\ \emph {et~al.}(2018)\citenamefont {Zhang},
  \citenamefont {Han}, \citenamefont {Wang}, \citenamefont {Car},\ and\
  \citenamefont {E}}]{deepmd_prl}%
  \BibitemOpen
  \bibfield  {author} {\bibinfo {author} {\bibnamefont {Zhang}, \bibfnamefont
  {L.}}, \bibinfo {author} {\bibnamefont {Han}, \bibfnamefont {J.}}, \bibinfo
  {author} {\bibnamefont {Wang}, \bibfnamefont {H.}}, \bibinfo {author}
  {\bibnamefont {Car}, \bibfnamefont {R.}}, and\ \bibinfo {author}
  {\bibnamefont {E}, \bibfnamefont {W.}},\ }\bibfield  {title} {\enquote
  {\bibinfo {title} {Deep potential molecular dynamics: A scalable model with
  the accuracy of quantum mechanics},}\ }\href
  {https://doi.org/10.1103/physrevlett.120.143001} {\bibfield  {journal}
  {\bibinfo  {journal} {Physical Review Letters}\ }\textbf {\bibinfo {volume}
  {120}} (\bibinfo {year} {2018}),\ 10.1103/physrevlett.120.143001}\BibitemShut
  {NoStop}%
\bibitem [{\citenamefont {Zhang}\ \emph {et~al.}(2014)\citenamefont {Zhang},
  \citenamefont {Ma}, \citenamefont {Fan}, \citenamefont {Zeng}, \citenamefont
  {Peng}, \citenamefont {Loya}, \citenamefont {Liu}, \citenamefont {Gong},
  \citenamefont {Zhang}, \citenamefont {Zhang}, \citenamefont {Ajayan},
  \citenamefont {Zhu},\ and\ \citenamefont {Lou}}]{nat_comm_KIC_smaller_value}%
  \BibitemOpen
  \bibfield  {author} {\bibinfo {author} {\bibnamefont {Zhang}, \bibfnamefont
  {P.}}, \bibinfo {author} {\bibnamefont {Ma}, \bibfnamefont {L.}}, \bibinfo
  {author} {\bibnamefont {Fan}, \bibfnamefont {F.}}, \bibinfo {author}
  {\bibnamefont {Zeng}, \bibfnamefont {Z.}}, \bibinfo {author} {\bibnamefont
  {Peng}, \bibfnamefont {C.}}, \bibinfo {author} {\bibnamefont {Loya},
  \bibfnamefont {P.~E.}}, \bibinfo {author} {\bibnamefont {Liu}, \bibfnamefont
  {Z.}}, \bibinfo {author} {\bibnamefont {Gong}, \bibfnamefont {Y.}}, \bibinfo
  {author} {\bibnamefont {Zhang}, \bibfnamefont {J.}}, \bibinfo {author}
  {\bibnamefont {Zhang}, \bibfnamefont {X.}}, \bibinfo {author} {\bibnamefont
  {Ajayan}, \bibfnamefont {P.~M.}}, \bibinfo {author} {\bibnamefont {Zhu},
  \bibfnamefont {T.}}, and\ \bibinfo {author} {\bibnamefont {Lou},
  \bibfnamefont {J.}},\ }\bibfield  {title} {\enquote {\bibinfo {title}
  {Fracture toughness of graphene},}\ }\href
  {https://doi.org/10.1038/ncomms4782} {\bibfield  {journal} {\bibinfo
  {journal} {Nature Communications}\ }\textbf {\bibinfo {volume} {5}} (\bibinfo
  {year} {2014}),\ 10.1038/ncomms4782}\BibitemShut {NoStop}%
\bibitem [{\citenamefont {Zhang}, \citenamefont {Li},\ and\ \citenamefont
  {Gao}(2015)}]{huajian_review_fracture}%
  \BibitemOpen
  \bibfield  {author} {\bibinfo {author} {\bibnamefont {Zhang}, \bibfnamefont
  {T.}}, \bibinfo {author} {\bibnamefont {Li}, \bibfnamefont {X.}}, and\
  \bibinfo {author} {\bibnamefont {Gao}, \bibfnamefont {H.}},\ }\bibfield
  {title} {\enquote {\bibinfo {title} {Fracture of graphene: a review},}\
  }\href@noop {} {\bibfield  {journal} {\bibinfo  {journal} {International
  Journal of Fracture}\ }\textbf {\bibinfo {volume} {196}},\ \bibinfo {pages}
  {1--31} (\bibinfo {year} {2015})}\BibitemShut {NoStop}%
\bibitem [{\citenamefont {Zhang}\ \emph {et~al.}(2019)\citenamefont {Zhang},
  \citenamefont {Zhang}, \citenamefont {Wang}, \citenamefont {Wang},
  \citenamefont {Zhang}, \citenamefont {Xu}, \citenamefont {Zou}, \citenamefont
  {Wu}, \citenamefont {Xia}, \citenamefont {Zhao},\ and\ \citenamefont
  {Wang}}]{brittle_exp_obser}%
  \BibitemOpen
  \bibfield  {author} {\bibinfo {author} {\bibnamefont {Zhang}, \bibfnamefont
  {Z.}}, \bibinfo {author} {\bibnamefont {Zhang}, \bibfnamefont {X.}}, \bibinfo
  {author} {\bibnamefont {Wang}, \bibfnamefont {Y.}}, \bibinfo {author}
  {\bibnamefont {Wang}, \bibfnamefont {Y.}}, \bibinfo {author} {\bibnamefont
  {Zhang}, \bibfnamefont {Y.}}, \bibinfo {author} {\bibnamefont {Xu},
  \bibfnamefont {C.}}, \bibinfo {author} {\bibnamefont {Zou}, \bibfnamefont
  {Z.}}, \bibinfo {author} {\bibnamefont {Wu}, \bibfnamefont {Z.}}, \bibinfo
  {author} {\bibnamefont {Xia}, \bibfnamefont {Y.}}, \bibinfo {author}
  {\bibnamefont {Zhao}, \bibfnamefont {P.}}, and\ \bibinfo {author}
  {\bibnamefont {Wang}, \bibfnamefont {H.~T.}},\ }\bibfield  {title} {\enquote
  {\bibinfo {title} {Crack propagation and fracture toughness of graphene
  probed by raman spectroscopy},}\ }\href
  {https://doi.org/10.1021/acsnano.9b03999} {\bibfield  {journal} {\bibinfo
  {journal} {{ACS} Nano}\ }\textbf {\bibinfo {volume} {13}},\ \bibinfo {pages}
  {10327--10332} (\bibinfo {year} {2019})}\BibitemShut {NoStop}%
\bibitem [{\citenamefont {Zhao}\ and\ \citenamefont {Aluru}(2010)}]{jap}%
  \BibitemOpen
  \bibfield  {author} {\bibinfo {author} {\bibnamefont {Zhao}, \bibfnamefont
  {H.}}and\ \bibinfo {author} {\bibnamefont {Aluru}, \bibfnamefont {N.~R.}},\
  }\bibfield  {title} {\enquote {\bibinfo {title} {Temperature and strain-rate
  dependent fracture strength of graphene},}\ }\href
  {https://doi.org/10.1063/1.3488620} {\bibfield  {journal} {\bibinfo
  {journal} {Journal of Applied Physics}\ }\textbf {\bibinfo {volume} {108}},\
  \bibinfo {pages} {064321} (\bibinfo {year} {2010})}\BibitemShut {NoStop}%
\bibitem [{\citenamefont {Zhao}\ \emph {et~al.}(2022)\citenamefont {Zhao},
  \citenamefont {Mao}, \citenamefont {Liu}, \citenamefont {Cao}, \citenamefont
  {Haigh}, \citenamefont {Papageorgiou}, \citenamefont {Li},\ and\
  \citenamefont {Young}}]{adv_mat_KIC}%
  \BibitemOpen
  \bibfield  {author} {\bibinfo {author} {\bibnamefont {Zhao}, \bibfnamefont
  {X.}}, \bibinfo {author} {\bibnamefont {Mao}, \bibfnamefont {B.}}, \bibinfo
  {author} {\bibnamefont {Liu}, \bibfnamefont {M.}}, \bibinfo {author}
  {\bibnamefont {Cao}, \bibfnamefont {J.}}, \bibinfo {author} {\bibnamefont
  {Haigh}, \bibfnamefont {S.~J.}}, \bibinfo {author} {\bibnamefont
  {Papageorgiou}, \bibfnamefont {D.~G.}}, \bibinfo {author} {\bibnamefont {Li},
  \bibfnamefont {Z.}}, and\ \bibinfo {author} {\bibnamefont {Young},
  \bibfnamefont {R.~J.}},\ }\bibfield  {title} {\enquote {\bibinfo {title}
  {Controlling and monitoring crack propagation in monolayer graphene single
  crystals},}\ }\href {https://doi.org/10.1002/adfm.202202373} {\bibfield
  {journal} {\bibinfo  {journal} {Advanced Functional Materials}\ ,\ \bibinfo
  {pages} {2202373}} (\bibinfo {year} {2022})}\BibitemShut {NoStop}%
\bibitem [{\citenamefont {Zhong}, \citenamefont {Li},\ and\ \citenamefont
  {Zhang}(2019)}]{jap_strain_hardening}%
  \BibitemOpen
  \bibfield  {author} {\bibinfo {author} {\bibnamefont {Zhong}, \bibfnamefont
  {T.}}, \bibinfo {author} {\bibnamefont {Li}, \bibfnamefont {J.}}, and\
  \bibinfo {author} {\bibnamefont {Zhang}, \bibfnamefont {K.}},\ }\bibfield
  {title} {\enquote {\bibinfo {title} {A molecular dynamics study of young's
  modulus of multilayer graphene},}\ }\href {https://doi.org/10.1063/1.5091753}
  {\bibfield  {journal} {\bibinfo  {journal} {Journal of Applied Physics}\
  }\textbf {\bibinfo {volume} {125}},\ \bibinfo {pages} {175110} (\bibinfo
  {year} {2019})}\BibitemShut {NoStop}%
\end{thebibliography}%

\end{document}